\pdfoutput=1
\documentclass[12pt]{article}
\usepackage{amssymb,lscape,psfrag}
\usepackage{epsfig,color}
\usepackage{array}
\usepackage{enumerate}
\usepackage{amsmath}
\usepackage{graphics}
\usepackage{epsfig}
\usepackage{verbatim}
\usepackage{cancel}
\usepackage{longtable}
\usepackage{multirow}
\usepackage{float}
\usepackage[normalem]{ulem} 
\usepackage[toc,page]{appendix}
\usepackage{rotating}
\usepackage{amsmath}
\usepackage{url}
\usepackage{bbm}    
\usepackage{amssymb}
\usepackage{amsmath}
\usepackage{graphicx}
\usepackage{hyperref}
\usepackage{natbib}
\usepackage{mathabx}
\usepackage{subfig}
\usepackage{scalerel}
\usepackage{xcolor}
\hypersetup{
	colorlinks,
	linkcolor={purple!80!black},
	citecolor={blue!50!black},
	urlcolor={black!100!black}
}
\usepackage{dsfont}
\usepackage{stmaryrd}

\newtheorem{theorem}{Theorem}[section]
\newtheorem{remark}{Remark}[section]

\newcommand {\bmath} {\begin {displaymath} }
\newcommand {\emath} {\end {displaymath} }

\newcommand {\by} {\mathbf{y}}
\newcommand {\diag} { \mbox{diag} }

\newcommand{\bb}{\boldsymbol{\beta}}

\newcommand{\bbk}{\boldsymbol{\beta}_k}

\newcommand{\Ip}{\mathbf{I}_{p}}

\newcommand{\X}{\mathbf{X}}
\newcommand{\XT}{\mathbf{X}^{T}}
\newcommand{\xxi}{\mathbf{x}_{i}}
\newcommand{\xxti}{\mathbf{x}_{i}^{T}}

\newcommand{\La}{\mathbf{S}}
\newcommand{\Laj}{\mathbf{S}_{k}}

\newcommand{\invLaj}{\mathbf{S}_{k}^{-1}}
\newcommand{\N}{\mathrm{N}}

\newcommand{\V}{\mathbf{V}}
\newcommand{\xii}{\mathbf{x}_{i}}

\newcommand{\M}{\mathbf{M}}
\newcommand{\sk}{\sigma^2_k}
\newcommand{\mk}{\boldsymbol{\mu}_k}
\newcommand{\Sk}{\boldsymbol{\Sigma}_k}
\newcommand{\s}{\boldsymbol{\sigma}^2}
\newcommand{\al}{\boldsymbol{\alpha}}
\newcommand{\m}{\boldsymbol{\mu}}
\newcommand{\Sig}{\boldsymbol{\Sigma}}

\newcommand{\lk}{\lambda_k}
\newcommand{\lj}{s_{kj}}
\newcommand{\mm}{\boldsymbol{\mu}}
\newcommand{\Om}{\mathbf{\Omega}}
\newcommand{\U}{\mathbf{U}_k^{(t)}}
\newcommand{\Usqrt}{\mathbf{U}_k^{\frac{1}{2}(t)}}
\newcommand{\f}{\boldsymbol{\phi}}

\newcommand{\thita}{\boldsymbol{\theta}_k}
\newcommand{\thitaY}{\boldsymbol{\theta}_k^{\scaleto{Y}{4pt}}}
\newcommand{\thitaX}{\boldsymbol{\theta}_k^{\scaleto{X}{4pt}}}
\newcommand{\ak}{\alpha_k}
\newcommand{\Ok}{\mathbf{\Omega}_k}
\newcommand{\thetab}{\boldsymbol{\theta}}
\newcommand{\xstar}{\mathbf{x}^*}

\DeclareMathOperator*{\argmin}{arg\,min}
\DeclareMathOperator*{\argmax}{arg\,max}

\newcommand{\R}{\mathbb{R}}

\newcommand{\parent}[1]{\left(#1\right)}

\renewcommand{\brace}[1]{\left\{#1\right\}}
\renewcommand{\det}[1]{\left|#1\right|}
\newcommand{\norm}[1]{\left\lVert#1\right\rVert}

\newcommand{\blind}{1}

\begin{document}

\def\spacingset#1{\renewcommand{\baselinestretch}%
{#1}\small\normalsize} \spacingset{1}

\date{}
\if1\blind
{
  \title{\bf Regularized joint mixture models}
  \author{Konstantinos Perrakis
\\
    \footnotesize{Department of Mathematical Sciences, Durham University, UK}\\
    Thomas Lartigue \\
    \footnotesize{Aramis Project Team, Inria \& Center of Applied Mathematics, CNRS, \'{E}cole Polytechnique, IP Paris, France}
    \\
    Frank Dondelinger \\
	\footnotesize{Lancaster Medical School, Lancaster University, UK}\\
    Sach Mukherjee \\
    \footnotesize{Statistics and Machine Learning, German Center for Neurodegenerative Diseases (DZNE), Bonn, Germany}}
  \maketitle
} \fi

\if0\blind
{
  \bigskip
  \bigskip
  \bigskip
  \begin{center}
    {\LARGE\bf Title}
\end{center}
  \medskip
} \fi

\begin{abstract}
	Regularized  regression models
	are well studied and, 
	under appropriate conditions,
	offer fast and statistically interpretable results.
	However, large data in many applications are 
	heterogeneous in the sense of harboring distributional differences between latent groups. Then,
	the assumption that the conditional distribution of response $Y$ given features $X$ 
	is the same for all samples may not hold.
	Furthermore, in scientific applications, the covariance structure of the features
	may contain important signals and 
	its learning is also affected by latent group structure.
	We propose a class of  mixture models for 
	paired data 
	$(X,Y)$ that 
	couples together 
	the  distribution of $X$ (using sparse graphical models)
	and the conditional $Y \mid X$ (using sparse regression models).
	The regression and graphical models are specific to the latent groups and model parameters are estimated jointly (hence the name ``regularized joint mixtures").
	This  allows
	signals in either or both of the feature distribution and regression model to inform learning of latent structure and 
	provides automatic control of confounding by such structure. Estimation is handled via an expectation-maximization algorithm, whose convergence is established theoretically. We illustrate the key ideas via empirical examples. An R package is available at \url{https://github.com/k-perrakis/regjmix}. 
\end{abstract}
\noindent%
\vfill

\newpage
\spacingset{1} 

\section{Introduction}
\label{intro}
Regularized linear models
usually assume homogeneity in the sense 
that the same 
conditional distribution  of a response $Y$ given features $X$ 
is taken to hold for all samples.
In the presence of latent groups that might have different underlying conditional distributions, regression modeling may be confounded, possibly severely. 
Similarly,  covariance structure among features can be an important signal in scientific applications but its learning may be strongly affected by latent group structure.

These issues are a concern whenever data might harbour unrecognized distributional shifts or group structure, an issue that is increasingly prominent in an era of large and often heterogeneous data. Furthermore, 
for heterogeneous data 
the two aspects -- the distribution of features $X$ and the conditional $Y | X$ -- are related in practice, since either or both may contain signals relevant to detecting and modeling group structure, which in turn is essential to overall estimation.
Motivated by such heterogeneous data settings with paired data of the form $(X,Y)$, in this paper we study a class of joint mixture models that couple together both aspects -- sparse graphical models for $X$ and parsimonious regression models for $Y|X$ -- in one framework. 
Specifically, in high-level notation, we consider models of the form
\begin{equation}
\begin{aligned}
Z  \sim & ~\tau_{\scaleto{Z}{4pt}} \\
X \mid Z {=} k  \sim & ~ p_{\scaleto{X}{4pt}}  (\mu_k,\Sigma_k) \\ 
Y \mid X, Z{=}k  \sim & ~p_{\scaleto{Y}{4pt}}(g(X^T \beta_k), \sigma_k^2)
\label{eq:model_intro}
\end{aligned}
\end{equation}
where $Z  \! \in \!  \{ 1,\dots,K \}$ is a latent indicator of group membership with  distribution  $\tau_{\scaleto{Z}{4pt}}$, $p_{\scaleto{W}{4pt}}(m,s)$ denotes the probability distribution of a 
random variable $W$ with location $m$ and scale $s$, and $g(\cdot)$ is a link function. This work  focuses on the familiar and important case where both $X$ and $Y$ are normally distributed and $g$ is the identity function. However, the key ideas apply to any model of the general form   in Eq. \eqref{eq:model_intro}.

In this context the presence of group structure has the following consequences:
\begin{itemize}
	\item {\it Confounding due to latent groups}. Associations between components of $X$ and $Y$ may be entirely different ``globally" (with $Z $ marginalized out) vs. ``locally"  (conditionally on $Z$) 
	with regression coefficients differing even in signs and sparsity patterns.

	\item {\it Ambiguous group structure in feature space}. Clustering the $X$'s alone 
	may lead to cluster labels which do not capture the relevant structure, as instances of group signal in $X$ 
	may be unrelated to $Y$ (e.g. clustering gene expression data may yield well-defined clusters; however, these may not relate to a specific biological/medical response).
	\item {\it Group-specific signal in regression coefficients}. 
	Nonidentical coefficients or feature importance across groups
	provide a potentially useful discriminant signal for identifying the group structure itself. 
	This signal 
	cannot be detected by clustering the $X$'s.
\end{itemize}

Two common 
strategies given paired data $(X,Y)$ 
are: 
(S1) ignore any potential grouping and fit one regression model
using the entire data 
and
(S2) cluster the $X$'s 
and then fit separate regression models to the group-specific data.
Strategy (S1) is particularly risky, since the resulting regression coefficients  
may be entirely incorrect 
if latent group structure is present
(e.g. due to Simpson's paradox and related phenomena).  
Also, when the modeling aspect is of importance,  
under (S1) evaluation of predictive loss is not a satisfactory guide for model assessment, since 
prediction error may be apparently small despite severe model misspecification. Strategy (S2) although safer is also not guaranteed to protect from such effects, unless the resulting group structure obtained from clustering the $X$'s is correct with reference to the overall problem; this may not hold in 
general.
Furthermore, since (S2) models the $X$ data alone, it cannot exploit any signal in the conditionals $Y|X$ to guide the clustering.
A common variant of (S2) is to perform 
a dimension reduction on $X$, such as PCA, and cluster on the reduced space. This, however, does not resolve the problem as the major principal components may not be predictive of $Y$;
see e.g. \cite{Jolliffe}.
\subsection{Related work and contribution}
For discrete latent variables $(Z)$ a standard way of approaching such heterogeneous problems is via mixture models. The literature on mixtures is vast; below we summarize related work according to model structure, covering the most popular approaches for the case of continuous responses.

{\it Mixtures for $X|Z$.}
The most commonly used and extensively studied approach within this category is the Gaussian mixture model (GMM); see e.g., \cite{McLachlan_Peel_2000}. GMMs have undergone a series of novel developments focusing on parsimonious modeling of the covariance matrices such as parametrizations based on eigenvalue decomposition \citep{banfield_raftery1993, celeux_govaert1995, fraley_raftery2002}, factorizations based on factor analysis models \citep{mcnicholas_murphy2008}, and extensions to sparse graphical model estimation  \citep{anandkumar2012,stadler2017,fop_etal2019}, among others.
These approaches differ from ours in that they consider only the $X$ signal and do not include a regression component, thus, inheriting the potential drawbacks of strategy (S2).

{\it Mixtures for $Y|X,Z$.} Finite mixtures of regression (FMR)
models belong in this category. Similarly to GMMs, Gaussian FMRs have been studied and developed extensively, allowing for flexible modeling designs \citep[see e.g.][]{fruwirth-schnatter_2005} and regularized estimation \citep{reqularizedFMR, Staedler_etal2010,khalili_lin2013}. 
FMRs focus on the relationship between $Y$ and $X$ without including a generative probability model for $X$. Our approach is motivated by settings in which the $X$ distribution itself is of interest and 
may be confounded by latent group structure.
Furthermore, under FMRs 
a {\it new} $X'$ cannot be allocated to {\it one specific} group and thereby used to obtain a group-specific prediction.

{\it Mixtures for $Y,Z|X$.} Mixtures of experts \citep[MoE;][]{dayton_macready1988,jacobs_etal1991,jordan1994,jacobs1997} model jointly the response and the latent allocations. MoEs consist of expert networks
(these are models that predict $Y$ from $X$) and a discriminative model (the gating network) that chooses among the experts.
The parsimonious covariance parametrizations for GMMs \citep{banfield_raftery1993, celeux_govaert1995, fraley_raftery2002} have been introduced within the MoE framework initially in \cite{dang_mcnicholas_2015} for the special case where the same set of predictors enter the expert and gating networks, and, more recently, in \cite{murphy_murphy2020} for the general case where different predictors are allowed to enter in the two networks; the latter work also introduces an additional noise component for outlier detection. Regularized MoE approaches include those of \cite{khalili2010} and \cite{chamroukhi2018}, among others. MoEs include FMRs as a special case in the absence of a gating network. Also, similarly to FMRs, MoEs condition upon features and, thus, lack a generative model for $X$. However, unlike with FMRs, group-specific prediction of the response is possible under MoEs, as the learned gating network can be used to allocate new feature observations. 

{\it Mixtures for $Y,X|Z$.} A first approach within this category is profile regression \citep{profile_regression, profile_regression1}.
Under profile regression $X$ and $Y$ are conditionally independent given the latent group indicator $Z$. Specifically, the component $Y|Z$ involves a regression model including a ``profile'' parameter (capturing the effect of $X$) plus additional co-variates, while the component $X|Z$ is some multivariate distribution (e.g. Gaussian). A second approach, more relevant to our work, is the cluster weighted model (CWM) mixture introduced by \cite{Ingrassia2012}. In this case, we have a  linear model component for $Y|X,Z$ and a multivariate distribution component for $X|Z$, following the hierarchical structure of Eq. \eqref{intro}. As illustrated, in \cite{Ingrassia2012} Gaussian CWMs lead to the same family of probability distributions generated by GMMs (more on that below) and under specific conditions include FMRs and MoEs as special cases. Extensions of CWMs accommodate the use of GLMs and mixed-type data \citep{ingrassia2015, punzo2016}, parsimonious parametrizations (similarly to GMMs and MoEs) of covariances \citep{dang_etal2017} and latent factor structures for the feature matrix \citep{subedi_etal2013}, among others.

The class of models proposed here -- henceforth, referred to as \textit{regularized joint mixture} (RJM) models -- belong to the latter category of mixtures. Specifically, the model specification (the likelihood part of the model) is of a CWM type, but the resulting clustering and parameter learning process under RJMs is different due to regularization.
CWMs rely on maximum-likelihood (ML) estimation and in this case 
under the normal-normal setting with identity link (as considered here), Eq. \eqref{eq:model_intro} is equivalent to a GMM on the concatenated matrix $[X,Y]$ as shown in \cite{Ingrassia2012}.
However, under RJMs, the equivalence to the GMM no longer holds, because the regression and graphical model parts are treated differently
(below we include 
comparisons with direct Gaussian mixture modeling of the concatenated matrix $[X,Y]$). We view regularization as essential for  delivering a usable solution to the problem.
In many cases the number of features $p$  may be on the same order as sample size $n$, or larger, and at the group level the sample sizes are of course smaller; hence 
without suitable regularization
both the regression and graphical models will typically be ill-behaved.
In the specific implementation we propose, we use the graphical lasso \citep{glasso2008} for graphical model estimation, while for the regression part we consider: (i) the Bayesian lasso  \citep{park_casella} 
and (ii) 
the normal-Jeffreys prior \citep{Figueiredo2001}. 
We note that other choices would be possible within the general framework, subject to computational considerations and appropriate handling of tuning parameters. 
In summary, the merits of RJMs are the following:
\begin{itemize}
	\item[](i)  Learns latent group structure by combining information from the distribution over $X$ and the regression of $Y$ on $X$
	within a principled  framework;
	\item[] (ii) 
	Provides group-specific feature importance  and graphical models with explicit sparsity patterns;
	\item[] (iii) Applicable in $p>n$ settings;
	\item[] (iv) Allows group-specific prediction for the response given a new feature vector $X'$.
\end{itemize}

\subsection{Motivation}
\label{motivation}
Given the large literature on mixture models, 
it is important to clarify at the outset why the models we study are needed.
We are motivated by applications in which 
latent group structure may be important and where
aspects such as  (potentially group specific) feature importance and covariance structure among the $X$'s play a role.

To take one example, in biomedicine there is much interest in  latent disease subtypes.  
These will often have subtype-specific covariance patterns, due to differences in underlying regulatory networks, and the analyst may want to understand subtype-specific disease biology and feature importance.
Focusing on only the $X$'s
may be insufficient, because this does not account for the response $Y$ (and  there may be many ways of clustering $X$ not relevant to the $Y$ of interest). 
For example, if $X$ are data on human subjects and $Y$ a cancer phenotype, many instances of cluster structure in $X$ may be unrelated to the cancer setting.
In addition, focusing solely on differences in regression models $Y | X$ means that subgroup recovery is difficult or impossible if these differences are not large enough. Similarly, in data-driven marketing, latent customer subgroups may have different covariance structure among features and at the same time manifest differences in regression models linking such features to responses (such as revenue per customer). The formulation we propose includes both sources of information in one model and thereby allows for subgroup identification and parameter estimation that accounts for both aspects. 

\begin{figure}[h]
	\centering{}\includegraphics[scale=0.6]{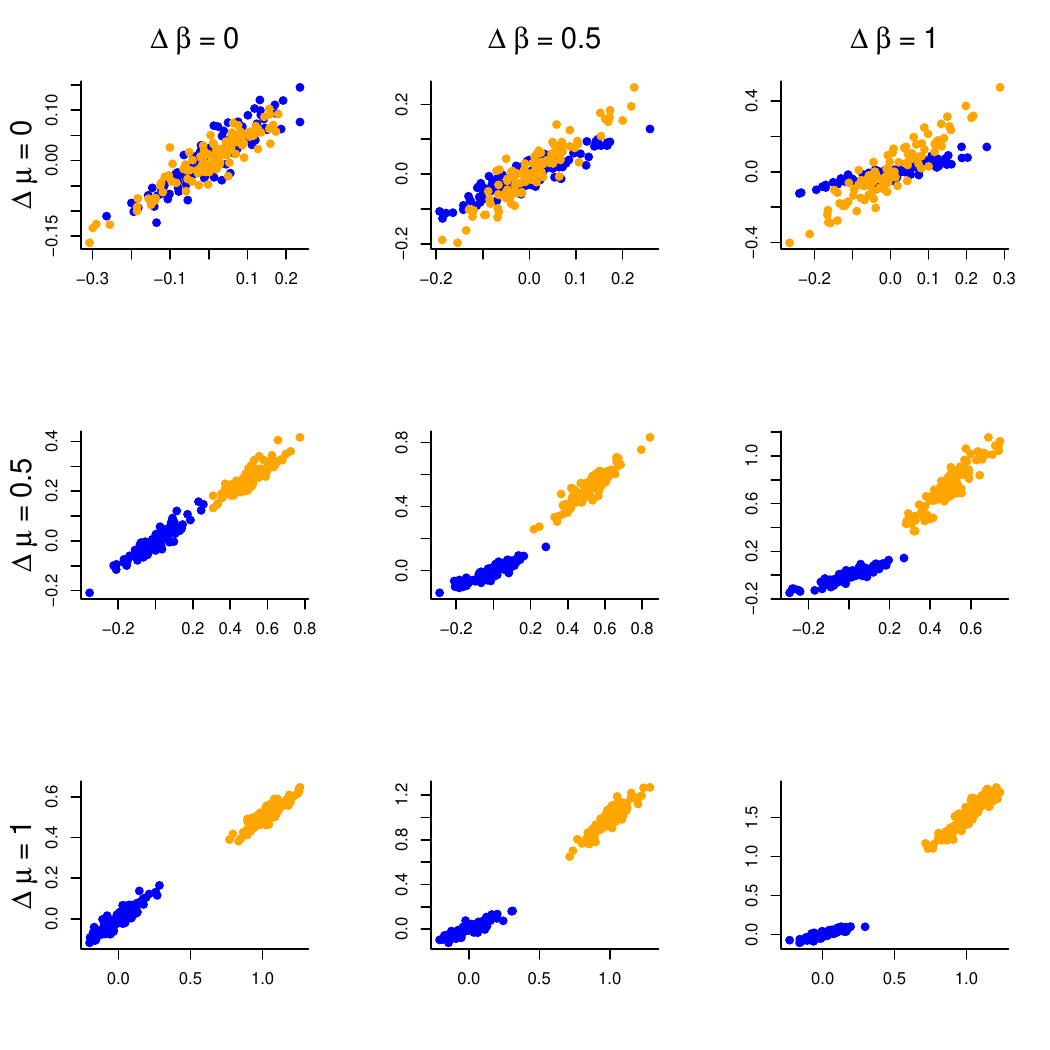}
	\caption{Examples for two subgroups. Each panel shows a specific level of difference in the regression models, quantified by the difference $\Delta \beta$ in regression coefficients. The difference in the feature distributions is controlled via a simple mean shift $\Delta \mu$.}
	\label{fig_intro1}
\end{figure}

\begin{figure}[h]
	\centering{}\includegraphics[scale=0.55]{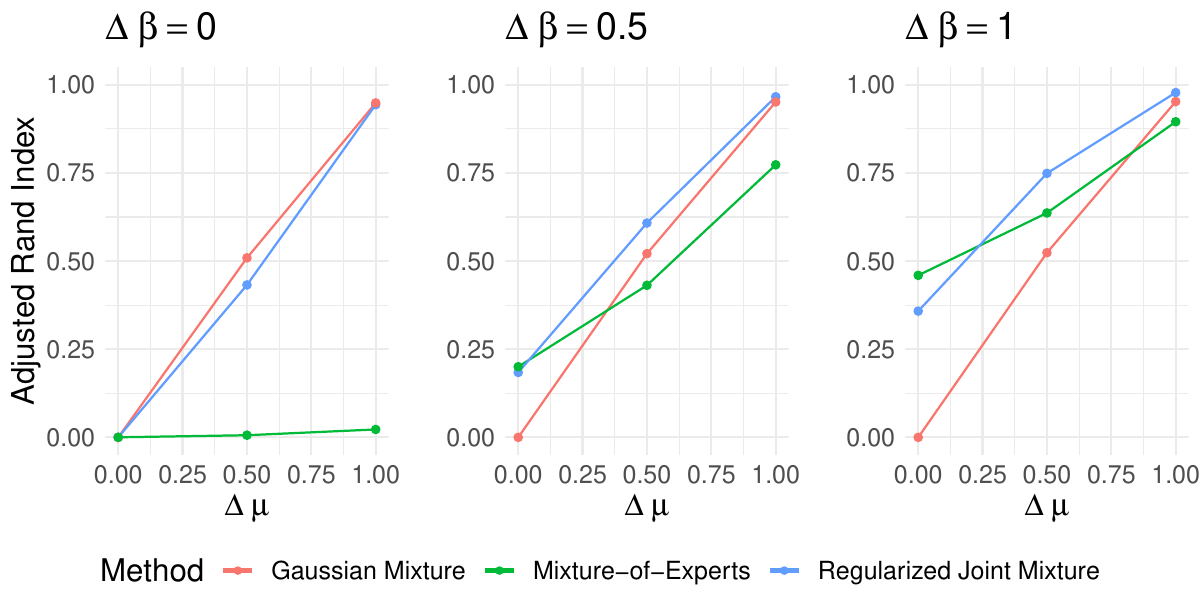}
	\caption{Role of signal location in subgroup identification. The difference $\Delta \beta$ in regression coefficients represents the signal from the  regression models, from no signal (left panel), to a strong signal (right panel). In each panel the signal in the feature distribution increases from left to right via a simple mean shift $\Delta \mu$. 
	}
	\label{fig_intro}
\end{figure}

Figure \ref{fig_intro1} shows simple illustrative examples to bring out some of these points. We emphasize that this is for illustrative purposes only, intended to highlight some interesting contrasts (full empirical results 
appear in Section \ref{simulations} below). In these examples, we consider settings in which there may be either or both of an $X$ signal difference (in the figure this is a simple mean shift $\Delta \mu$, but the models we propose are general and 
for multivariate $X$ 
allow also for differences in covariance structure) and a difference in subgroup-specific regression coefficients ($\Delta \beta$).
For this initial illustration we consider two latent groups, each with sample size equal to 100, and ten potential predictors, but with only one predictor having an effect (a non-zero regression coefficient) on the response; analytic details of the simulations can be found in Appendix A.
Results in terms of subgroup identification, as quantified by Rand Index, are summarized in Figure \ref{fig_intro}. 
When there is no difference in regression models ($\Delta \beta = 0$), MoEs cannot detect any structure (since the $X$ distribution is not modelled). On the other hand, with a stronger difference in regression models ($\Delta \beta = 1$), MoE outperforms a Gaussian mixture (on the $X$'s), since the latter does not model the regression part. The approach we propose models both aspects in a unified framework, hence works well regardless of where the signal lies.
Furthermore, and as shown in detail via empirical examples below, by accounting for the latent structure, RJM is able to detect subgroup-specific sparsity patterns whilst avoiding Simpson's paradox-like effects that could otherwise arise. 
Later, we show detailed empirical results, including an example, based on cancer data, that highlights  some of these points, and in particular how subgroup identification benefits from joint modeling, relative to simply clustering $X$ (or clustering the stacked vector $(X, Y)$) or using MoE. 

The remainder of the paper is structured as follows. 
In Section \ref{model} we lay out the model specification and discuss the regularization methods under consideration as well as efficient tuning strategies.
Computational and theoretical details of the expectation-maximization (EM) optimization are covered in Section \ref{EM}. 
In Section \ref{prediction} we discuss prediction using RJMs and discuss how predictive measures can potentially be used for cluster selection.
In Section \ref{simulations} we present empirical examples, focusing initially on small-scale 
simulations and then proceeding to larger scale semi-synthetic experiments and applications to real data. The paper concludes with a discussion in Section \ref{disc}. 

\section{The RJM model}
\label{model}


\subsection{Model specification}
\label{likelihood}
Let $\by$ denote an $n$-dimensional vector of outputs or responses  
and $\X$ an  $n \times  p$ feature matrix. 
Samples are indexed by $i = 1,\ldots, n$.
Let $K$ denote the number of groups and $z_i \in \{1 \ldots K \}$ represent the true (latent) group indicator for the sample point $(y_i,\xii)$ with $\Pr(z_i=k)=\tau_k$.
The group-specific parameters are $\thita = (\thitaX,\thitaY)^T$ with $\thitaX$ and  $\thitaY$ being the parameters  governing respectively the marginal distribution of $X$ and the regression model of $Y$ on $X$.

We allow for group-specific parameters, but assume that samples are independent and identically distributed within groups. 
The joint distribution of $(y_i,\xii)$ in group $k$ is 
\begin{equation}
p(y_i,\xxi|\thita,z_i=k)=p(y_i|\thitaY,\xxi,z_i=k)p(\xxi|\thitaX,z_i=k).
\label{lik_joint}
\end{equation}
The features are modeled as $p$-dimensional multivariate normal so that $\thitaX=(\mk,\mathrm{vec}(\Sk))^T$, where $\mk$ is the mean and $\Sk$ the $p\times p$ covariance matrix.
For the responses, we specify a normal linear regression model, with parameters $\thitaY=(\ak,\bbk,\sk)^T$, where $\ak$ is the intercept, $\bbk$ the vector of regression coefficients and $\sk$ the error variance.  Inclusion of the intercept is necessary in the present setting, because it is not possible to center  the response appropriately when the group labels are unknown. 
Thus, we have that  
\begin{equation}
p(\xxi|\thitaX,z_i=k) \equiv p(\xxi|\mk,\Sk,z_i=k) = \mathrm{N}_p(\xxi|\mk,\Sk)
\label{lik_x}
\end{equation}
and
\begin{equation}
p(y_i|\thitaY,\xxi,z_i=k) \equiv p(y_i|\ak,\bbk,\sk,\xxi,z_i=k) = \mathrm{N}(y_i|\ak+\xxti\bbk,\sk).
\label{lik_y}
\end{equation}
Marginalizing out the latent variables leads to a mixture representation of the form
\begin{equation}
p(\by,\X|\boldsymbol{\theta},\boldsymbol{\tau})=\prod_{i=1}^{n}\sum_{k=1}^{K}
p(y_i|\thitaY,\xxi,z_i=k)p(\xxi|\thitaX,z_i=k)\tau_k,
\label{lik_marg}
\end{equation}
where $\boldsymbol{\theta}=(\boldsymbol{\theta}_1,\dots,\boldsymbol{\theta}_K)^T$ and $\boldsymbol{\tau}=(\tau_1,\dots,\tau_K)^T$.

\subsection{Regularization and priors}
\label{penalization}

Given the likelihood function of $\boldsymbol{\theta}$ and $\boldsymbol{\tau}$ in \eqref{lik_marg} we consider general solutions of the form
\begin{equation}
\boldsymbol{\hat{\theta}}, \boldsymbol{\hat{\tau}} = \underset{\boldsymbol{\theta},\boldsymbol{\tau}}{\argmax} \Bigg\{ \log p(\by,\X|\boldsymbol{\theta},\boldsymbol{\tau})+\sum_{k=1}^{K}\mathrm{pen}(\thita^{*{\scaleto{X}{4pt}}})+\sum_{k=1}^{K}\mathrm{pen}(\thita^{*{\scaleto{Y}{4pt}}})\Bigg\},
\label{pen_est}
\end{equation} 
where $\mathrm{pen}(\cdot)$ denotes a penalty function,
and $\thita^{*{\scaleto{X}{4pt}}}$ and $\thita^{*{\scaleto{Y}{4pt}}}$
are parameter subsets 
that we wish to penalize. 
The particular parameters we penalize  are the group-specific  covariances   and 
regression vectors; hence,    $\thita^{*{\scaleto{X}{4pt}}}\equiv\mathrm{vec}(\Sk)$ and $\thita^{*{\scaleto{Y}{4pt}}}\equiv\bbk$ in  \eqref{pen_est} (although under one approach below we consider $\thita^{*{\scaleto{Y}{4pt}}}\equiv(\bbk,\sk)^T$). 
Penalization is required because the corresponding ML estimates may be  ill-behaved or ill-defined
and we are interested in group-specific feature importance and  conditional independence structure.


In general, tuning of penalty parameters is challenging in the latent group setting. The common approach, based on cross-validation (CV), needs to be handled with extreme care when the estimation of parameters requires iterative procedures converging to (local) maxima, such as the EM algorithm. Specifically, performing CV at each iteration of the algorithm would change the penalty and, thus, also change the objective function at each iteration. A brute-force solution to the problem would be to pre-specify a grid of values for the penalty and select the value that optimizes a specific criterion. However, apart from open issues related to the range and length of the grid, this would also be a computationally burdensome task, requiring multiple EM processes (each with multiple starts) for each point at the grid. 

Given these considerations, we believe that more viable strategies are the following; (i) using ``universal penalties'' from existing literature, (ii) using CV in a stepwise manner, and (iii)  considering the penalties as free parameters under estimation. The first strategy is advisable to use for parameters whose learning is not the main goal of the analysis, but for which regularization is required in order to attain workable, non-spurious solutions.  Universal penalties which satisfy certain theoretical requirements \citep[see e.g.,][]{donoho1994, Staedler_Mukherjee2013} typically work well to that end. On the other hand, for the main parameters of interest, whose learning (e.g. estimation, sparsity patterns) is of importance, the second and third strategies seem to be more appropriate as they offer more specific, application-driven solutions. In short, the stepwise CV approach entails adjusting the penalties a few times during the EM and running the algorithm sufficiently long in order to reach local maxima after the last adjustment. The last strategy, requires maximizing the objective with respect to the penalties and, thus, requires the introduction of a prior distribution. Within this framework one could ideally (yet not necessarily) consider properties of the prior; for instance, its behavior under small and/or large samples and the consequent effect on shrinkage, among others. From a pragmatic perspective, the prior choice is linked to more realistic considerations relating to the maximization required in the EM algorithm. For instance, the half-Cauchy prior, which is a standard choice for penalties \citep{Polson_Scott_2012} in fully Bayesian implementations based on posterior sampling, may not be necessarily convenient to work with within an EM framework.

We proceed with a description of the 
penalty functions, discussing our choices from both penalized likelihood and Bayesian viewpoints, and explaining our reasoning for the way that penalty parameters are tuned based on the aforementioned strategies.
For the remainder of this Section, 
it is convenient to discuss the corresponding solutions under known group labels. For the subset of datapoints where $z_i=k$, let us denote the $n_k \times 1$ response by $\by_k$ and the $n_k \times p$ predictor matrix by $\X_k$ for $k = 1,\dots,K$. 
We emphasize that this is for expositional clarity only; the 
actual solutions under latent group labels are, of course, obtained iteratively via the EM algorithm presented in Section \ref{EM}.

\subsubsection{Regularization of $\Sk$}
\label{glasso}

For the regularization of $\Sk$ we use the graphical lasso \citep{Meinshausen2006, Yuan2007, glasso2008}. The graphical lasso induces sparsity in the inverse covariance matrix
$\Ok=\Sk^{-1}$ for group $k$. In this case we have  $\mathrm{pen}(\thita^{*{\scaleto{X}{4pt}}})\equiv\mathrm{pen}(\Ok)=-\zeta\Vert\Ok\Vert_1$ in \eqref{pen_est}, where  $\zeta>0$ controls the strength  of regularization and $\Vert\cdot\Vert_q$ is the $L_q$ norm. For known group labels the graphical lasso estimate would be
\begin{equation}
\widehat{\boldsymbol{\Omega}}_k=\underset{\Ok\in M^+}{\argmax}\Big\{\log |\Ok|  - \mathrm{tr}\big(\Ok\hat{\mathbf{S}}_k\big) -\zeta\Vert\Ok\Vert_1\Big\},
\label{glasso_pen}
\end{equation}    
where $M^+$ is the space of positive definite matrices and $\hat{\mathbf{S}}_k$ is the ML covariance estimate of $\X_k$.
The solution in \eqref{glasso_pen} is equivalent to the posterior mode under a likelihood as in \eqref{lik_x} and a prior distribution of the following form    
\begin{equation}
p(\Ok|\psi) \propto 
\Bigg[
\prod_{j=1}^p\mathrm{Exp}(\omega_{kjj}|\psi/2)
\prod_{j<l,l=2}^p
\mathrm{DE}(\omega_{kjl}|0,\psi^{-1})
\Bigg]\mathbbm{1}_{\{\Ok\in M^+\}},
\label{glasso_prior}
\end{equation}
where $\mathbbm{1}_{\{\cdot\}}$ is the indicator function, $\mathrm{Exp}(\cdot|r)$ is the exponential distribution with rate $r>0$ and $\mathrm{DE}(\cdot|\mu,b)$ is the double exponential distribution with location $\mu\in\mathbb{R}$ and scale $b>0$. 
The connection between \eqref{glasso_pen} and the corresponding posterior mode under the likelihood in \eqref{lik_x} and the prior in \eqref{glasso_prior} is that for any given value of $\zeta$ we have that $\psi=n_k\zeta$ \citep{Wang2012}.

For the graphical lasso penalty, we use the ``universal" threshold  $\tilde{\psi}=\sqrt{2n\log p}/2$ derived from the logical arguments developed in \cite{Staedler_Mukherjee2013}. Note that here we use $n$ instead of $n_k$ as the group labels will be unknown.  The reason for choosing the universal-threshold approach for the graphical lasso 
is that we are less interested in the sparsity pattern of $\Ok$ itself. 
Rather, we mainly want a well-behaved estimate that allows group structure to be effectively accounted for.

\subsubsection{Regularization of $\bbk$}
\label{regression}


\textbf{The lasso approach.} We consider a scaled version of the lasso \citep{tibshirani}, where the penalty in \eqref{pen_est} is given by $\mathrm{pen}(\thita^{*{\scaleto{Y}{4pt}}})\equiv\mathrm{pen}(\bbk,\sk)=\lk\Vert\bbk\Vert_1/\sigma_k$ with $\lk >0$. 
Here we introduce group specific penalty parameters $\lk$, unlike previously, where parameter $\psi$ is common across groups. This type of lasso regularization has been studied by \cite{Staedler_etal2010} in the context of FMR; a setting where the role of the scaling of variance parameters is much more important than in standard homogeneous regression, as the $\sigma_k$'s may differ and the grouping is not fixed.
If the groups were known, the lasso estimate would be
\begin{equation}
\hat{\alpha}_k,\hat{\bb}_k,\hat{\sigma}^2_k=\underset{\ak,\bbk,\sk}{\argmin}\Bigg\{\frac{\Vert \by_k-\ak\mathbf{1}_{n_k}-\X_k\bbk\Vert_2^2}{2\sk}+\lk\frac{\Vert\bbk\Vert_1}{\sigma_k}+(n_k+p+2)\log\sigma_k\Bigg\},
\label{lasso_sol}
\end{equation}
where $\mathbf{1}_q$ denotes a $q$-dimensional vector of ones. 
The solution in \eqref{lasso_sol}, which is slightly different than the one in \cite{Staedler_etal2010}, corresponds  to the posterior mode under the Bayesian lasso formulation \citep{park_casella}, that specifies independent double exponential priors for the regression coefficients conditional on the error variance which is assigned the scale-invariant Jeffreys prior. Namely,  
\begin{equation}
p(\bbk|\sk,\lk)=\displaystyle \prod_{j=1}^p 
\frac{\lk}{2\sigma_k}
\exp\Bigg(-\lk\displaystyle\frac{\vert\beta_{kj}\vert}{\sigma_k}\Bigg) \mbox{~ and ~} p(\sk)\propto \frac{1}{\sk},
\label{lasso_prior}
\end{equation} 
with the correspondence to \eqref{lasso_sol} completed when $p(\alpha_k)\propto 1$.
We propose two methods for handling $\lk$; the {\it fixed-penalty lasso} (FLasso) based on a  plug-in estimate and the {\it random-penalty lasso} (RLasso) based on the construction of a suitable prior.

\medskip
\noindent 
\textit{FLasso.} This approach is essentially a two-step tuning procedure. We start with initial estimates, $\hat{\lambda}_k^{\scaleto{(0)}{6pt}}$ obtained by minimizing the CV mean squared error based on some prior clustering of the data. 
Then, at a certain iteration we re-calculate the CV estimates and fix each group penalty to the new estimate $\hat{\lambda}_k^{\scaleto{(1)}{6pt}}$ for all further EM iterations. Specifically, we fix the parameter after the first iteration where the group assignments 
do not change. From a Bayesian perspective the FLasso approach can be viewed as an empirical Bayes method as we use the data in order to plug-in $\hat{\lambda}_k^{\scaleto{(0)}{6pt}}$ and $\hat{\lambda}_k^{\scaleto{(1)}{6pt}}$ in the prior of $\bbk$ appearing in \eqref{lasso_prior}. The monotonic behaviour of the EM may be disrupted at the re-estimation iteration, but after that point it will hold. 

\medskip

\noindent 
\textit{RLasso.} In this approach we propose placing a prior distribution on $\lambda_k$ so that this parameter will be automatically updated during the EM.
We construct the prior so that it satisfies the requirement of supporting no penalization asymptotically ($\lambda_k\rightarrow0$ as $n\rightarrow\infty$).
A suitable prior for our purposes is the Pareto distribution whose scale parameter is also the lower bound of its support. 
Specifically, we have a prior distribution with scale $a_n>0$ and shape $b_n>0$ (parameters are defined to depend on $n$) of the following form 
\begin{equation}
p(\lambda_k)=b_n a_n^{b_n}\lambda_k^{-(b_n+1)}, 
\label{prior_lambda}
\end{equation}
where $\lambda_k\in[a_n, \infty)$.
In our setting parameter $a_n$ does need to be specified explicitly and is regarded to be decreasing in $n$, while the shape parameter is specified as $b_n=(p-1)- c\sqrt{2K\log p/n}$ for some $c\in (0,1]$. The rationale for these choices and further details are discussed in Appendix B.     
As shown next, RLasso will lead to a reasonable  update for $\lambda_k$ during the M-step. 
\\  
\\
\noindent \textbf{The normal-Jeffreys approach.}
The normal-Jeffreys (NJ) prior \citep{Figueiredo2001} consists of independent improper priors; 
in our context the prior is given by
\begin{equation}
p(\bbk) = \prod_{j=1}^p p(\beta_{kj}) ~ \propto ~ \prod_{j=1}^p|\beta_{kj}|^{-1}.
\label{nj}
\end{equation}
For known group labels with $p(\ak, \sk) \, \propto \, 1/\sk$ the corresponding penalized estimate is 
\begin{equation}
\hat{\alpha}_k,\hat{\bb}_k,\hat{\sigma}^2_k=\underset{\ak,\bbk,\sk}{\argmin}\Bigg\{\frac{\Vert \by_k-\ak\mathbf{1}_{n_k}-\X_k\bbk\Vert_2^2}{2\sk}+\sum_{j=1}^p\log|\beta_{kj}|+(n_k+2)\log\sigma_k\Bigg\}.
\label{nj_sol}
\end{equation}
Here 
$\mathrm{pen}(\thita^{*{\scaleto{Y}{4pt}}})\equiv\mathrm{pen}(\bbk)=\sum_{j=1}^p\log|\beta_{kj}|$.
The NJ is well known in the 
shrinkage-prior literature \citep{Griffin_Brown2005,carvalho_etal_2010,Polson_Scott_2010}. As with most shrinkage priors,
\eqref{nj} can be expressed as a scale-mixture of normals;  
namely,
$
p(\beta_{kj}|s_{kj}) = (2\pi s_{kj})^{-1/2}\exp(-\beta_{kj}^2/2s_{kj})
\label{scale_mixture}
$
with $\pi(s_{kj}) \, \propto \, s_{kj}^{-1}$ and, therefore, we have that $\int p(\beta_{kj}|s_{kj}) \pi(s_{kj})\mathrm{d}s_{kj} \, \propto \, |\beta_{kj}|^{-1} $. 
As the mixing distribution lacks a 
hyper-parameter the prior is characterized by the absence of a ``global'' scale parameter. 
Also, due to heavy tails 
small coefficients are shrunk a lot, while large signals remain relatively unaffected; similarly to other heavy-tailed priors  \citep{carvalho_etal_2010,Griffin_Brown2005}.
\cite{Figueiredo2003} and \cite{BaeMallick2004} show that the NJ prior strongly induces sparsity and yields good performance in terms of selection. 

The use of the NJ prior is appealing for the RJM framework.
Handling penalties is cumbersome in our setting 
and the NJ prior provides an attractive  ``tuning-free" alternative. 
In general, shrinkage priors which lack a global scale parameter fail to capture the average signal density of the data \citep{carvalho_etal_2010}; however, despite this potential shortcoming of the NJ prior the potential benefits are worth exploring. Also, the posterior mode under \eqref{nj} is easy to find through the use of an EM algorithm where the scaling parameters $s_{kj}$ are considered latent \citep{Figueiredo2003}. This additional latent structure can be easily incorporated in our EM without additional computational costs. In fact, the corresponding NJ-EM update is in closed-form, which is not the case in the lasso approach.

\section{The RJM-EM algorithm}
\label{EM}

In this Section we present first the expectation and maximization steps of the proposed EM algorithm. We then prove that under certain conditions on the regularization the proposed algorithm converges towards a critical point of the likelihood function. 

\subsection{The EM steps}
\label{steps}

\textbf{The E-Step.}  Irrespective of regularization approach, the group-membership probabilities of the mixture model in \eqref{lik_marg} at iteration $t$ of the algorithm are calculated as 
\begin{equation}
m^{(t)}_{ki}\equiv\widehat{\Pr}(z_i=k|y_i,\xxi,\thita^{(t)}) = 
\frac{p(y_i|\thita^{{\scaleto{Y}{4pt}}(t)},\xxi,z_i=k)p(\xxi|\thita^{{\scaleto{X}{4pt}}(t)},z_i=k)\tau_k^{(t)}}
{\sum_kp(y_i|\thita^{{\scaleto{Y}{4pt}}(t)},\xxi,z_i=k)p(\xxi|\thita^{{\scaleto{X}{4pt}}(t)},z_i=k)\tau_k^{(t)}},
\label{member}
\end{equation}
for $i=1,\dots,n$, with the distributions appearing in the right-hand side of \eqref{member} defined in \eqref{lik_x} and \eqref{lik_y}. Let us define some quantities that will be used throughout; namely, $ 
n_k^{(t)}=\sum_{i=1}^{n}m_{ki}^{(t)}$,  $\mathbf{m}_k^{(t)}=(m_{k1}^{(t)},\dots,m_{kn}^{(t)})^T$ and  $\M_k^{(t)}=\diag(\mathbf{m}_k^{(t)})$.

A convenient feature of the RJM design is that due to the hierarchical structure of the model the objective function can be split into separate simple parts; specifically, 
\begin{equation}
Q(\thetab,\boldsymbol{\tau},\boldsymbol{\lambda}|\thetab^{(t)},\boldsymbol{\tau}^{(t)},\boldsymbol{\lambda}^{(t)})=
Q^{\scaleto{Y}{4pt}}(\thetab^{\scaleto{Y}{4pt}},\boldsymbol{\lambda}|\thetab^{{\scaleto{Y}{4pt}}(t)},\boldsymbol{\lambda}^{(t)})
+Q^{\scaleto{X}{4pt}}(\thetab^{\scaleto{X}{4pt}}|\thetab^{{\scaleto{X}{4pt}}(t)})+
Q^{\scaleto{Z}{4pt}}(\boldsymbol{\tau}|\boldsymbol{\tau}^{(t)}),
\label{obj}
\end{equation}      
where $\thetab^{\scaleto{Y}{4pt}}=(\thetab^{\scaleto{Y}{4pt}}_1,\dots,\thetab^{\scaleto{Y}{4pt}}_K)^T$ and $\thetab^{\scaleto{X}{4pt}}=(\thetab^{\scaleto{X}{4pt}}_1,\dots,\thetab^{\scaleto{X}{4pt}}_K)^T$. 
Here, by $\boldsymbol{\lambda}$ we denote the vector of group penalty parameters of the regression component. Depending upon regularization approach, the elements of vector $\boldsymbol{\lambda}$ at iteration $t$ are fixed in  FLasso, free and under estimation in RLasso and absent in  NJ; 
respectively having $\boldsymbol{\lambda}^{(t)}=(\hat{\lambda}_1 ^{(t^*)},\dots,\hat{\lambda}_K ^{(t^*)})^T$
(where $t^*=\{0,1\}$ with zero and one corresponding to the initial CV estimate and the re-estimated CV value; see FLasso in Section \ref{regression}),
$\boldsymbol{\lambda}^{(t)}=({\lambda}_1 ^{(t)},\dots,{\lambda}_K ^{(t)})^T$ and
$\boldsymbol{\lambda}^{(t)}=\varnothing$.

Starting in reverse  order from the right-hand side of \eqref{obj} we have that
\begin{equation}
Q^{\scaleto{Z}{4pt}}(\boldsymbol{\tau}|\boldsymbol{\tau}^{(t)})=\sum_{k=1}^K n_k^{(t)}\log\tau_k,
\label{obj_z}
\end{equation}
while the second component of the objective function is given by     
\begin{equation}
Q^{\scaleto{X}{4pt}}(\thetab^{\scaleto{X}{4pt}}|\thetab^{{\scaleto{X}{4pt}}(t)}) = 
\frac{1}{2}
\sum_{k=1}^{K}\Bigg[
\sum_{i=1}^{n} m^{(t)}_{ki} \Bigg[\log|\Om_k|-(\xxi-\mk)^{T}\Om_k(\xxi-\mk)\Bigg]
-\tilde{\psi}\Vert\Om_k\Vert_1
\Bigg],
\label{obj_x}
\end{equation}
where $\tilde{\psi}=\sqrt{2n\log p}/2$.
The last component $Q^{\scaleto{Y}{4pt}}$ in \eqref{obj} depends on  regularization method. 
We define two distinct functions $Q_{\scaleto{\mathrm{lasso}}{4pt}}^{\scaleto{Y}{4pt}}$ and $Q_{\scaleto{\mathrm{NJ}}{4pt}}^{\scaleto{Y}{4pt}}$. For lasso we use the re-parametrization $\chi_k=\ak/\sigma_k$, $\f_k=\bbk/\sigma_k$ and $\rho_k=\sigma_k^{-1}$ \citep{Staedler_etal2010}, resulting in
\begin{align}
Q^{\scaleto{Y}{4pt}}_{\scaleto{\mathrm{lasso}}{4pt}}(\thetab^{\scaleto{Y}{4pt}},\boldsymbol{\lambda}|\thetab^{{\scaleto{Y}{4pt}}(t)},\boldsymbol{\lambda}^{(t)}) = &
\sum_{k=1}^{K}  \displaystyle\Bigg[ 
-\frac{(\rho_k\by-\chi_k\mathbf{1}_{n}-\X\f_k)^{T}\M_k^{(t)}(\rho_k\by-\chi_k\mathbf{1}_{n}-\X\f_k)}{2} \nonumber \\
&
\qquad-\lambda_{k}\Vert\f_k\Vert_1 +(n_k^{(t)}+p+2)\log\rho_k 
+f(\lambda_k)\Bigg],
\label{obj_y_lasso}
\end{align}
which is convex. Here $f^{(t)}(\lambda_k)=0$ for FLasso and 
$f(\lambda_k)=c\displaystyle\sqrt{2K\log p/n}\log \lambda_k $ for RLasso.
Under the NJ approach the corresponding objective is given by
\begin{align}
Q^{\scaleto{Y}{4pt}}_{\scaleto{\mathrm{NJ}}{4pt}}(\boldsymbol{\theta}^{\scaleto{Y}{4pt}}|\boldsymbol{\theta}^{{\scaleto{Y}{4pt}}(t)}) = & 
-\frac{1}{2}\sum_{k=1}^{K}  \Bigg[ 
\frac{(\by-\ak\mathbf{1}_{n}-\X\bbk)^{T}\M_k^{(t)}(\by-\ak\mathbf{1}_{n}-\X\bbk)}{\sk}+\bbk^{T}\V^{(t)}_k\bbk \nonumber
\\
&
\qquad \qquad+(n_k^{(t)}+2)\log\sk
\Bigg],
\label{NJobj}
\end{align}
where 
$
\V_k^{(t)} = \diag\Big(1/\beta_{k1}^{2(t)},\dots,1/\beta_{kp}^{2(t)}\Big)
$.
This matrix arises from the second underlying latent structure (the latent scale parameters) in the EM;
details are provided in Appendix C.  
As we will see below, matrix $\V_k$ will in fact provide the final sparse estimate of $\bb_k$, as some of its diagonal entries go to infinity during the EM; consequently, the diagonal entries of $\mathbf{U}_k=\V_k^{-1}$ that go to zero correspond to the coefficients that are set equal to zero.

\medskip
\noindent\textbf{The M-Step.} From \eqref{obj_z} we have that the group probabilities are updated as 
\begin{equation}
\tau^{(t+1)}_k {=}n_k^{(t)}/n.
\label{tau}
\end{equation}
Concerning parameter block $\thitaX$, from \eqref{obj_x} we obtain the following updating equations
\begin{equation}
\boldsymbol{\mu}_k^{(t+1)} = \displaystyle\frac{\sum_{i=1}^n m_{ki}^{(t)} \mathbf{x}_i}{n_k^{(t)}},
\label{mu}
\end{equation}
\begin{equation}
\Om_k^{(t+1)}= \underset{\Om_k}{\argmax}\Big\{\log |\Om_k|  - \mathrm{tr}\big(\Om_k\mathbf{S}_k^{(t)}\big) -\zeta_k^{(t)}\Vert\Om_k\Vert_1\Big\},
\label{omega}
\end{equation}
where in \eqref{omega} we have that $\mathbf{S}_k^{(t)} = n_k^{-(t)}\sum_{i=1}^n m_{ki}^{(t)} (\mathbf{x}_i - \boldsymbol{\mu}_k^{(t+1)}) (\mathbf{x}_i - \boldsymbol{\mu}_k^{(t+1)})^T $ and penalty given by $\displaystyle \zeta_k^{(t)}=\tilde{\psi}/n_k^{(t)}=\sqrt{2n\log p}/(2n_k^{(t)})$. 
For the lasso objective in \eqref{obj_y_lasso} the FLasso group penalties are simply $\lambda_{k}^{(t+1)}=
\hat{\lambda}_k^{(t^*)}$, while the RLasso update is 
\begin{equation}
\lambda_{k}^{(t+1)}=
\displaystyle\frac{cK^{1/2}}{\Vert\f_k^{(t)}\Vert_1}\sqrt{\frac{2\log p}{n}}
= \frac{cK^{1/2}}{\Vert\bbk^{(t)}\Vert_1}\Bigg( \sigma_k^{(t)}\sqrt{\frac{2\log p}{n}} \Bigg).
\label{lambda}
\end{equation} 
Note that the RLasso update is a scaled version of the optimal universal penalty under orthonormal predictors (quantity inside the parenthesis in  \eqref{lambda}) and that the scaling depends on the sparsity of the coefficients and on $c$.
For the components of $\thitaY$, the updates are as follows
\begin{equation}
\rho_k^{(t+1)} = \displaystyle\frac{\by^T \M_k^{(t)} (\chi_k^{(t)}\mathbf{1}_{n}+\X\f_k^{(t)}) + \sqrt{ \parent{\by^T \M_k^{(t)} (\chi_k^{(t)}\mathbf{1}_{n}+\X\f_k^{(t)})}^2 + 4 \by^T \M_k^{(t)} \by (n_k^{(t)}+p+2) }}{2 \by^T \M_k^{(t)} \by },
\label{rho}
\end{equation}
\begin{equation}
\chi_k^{(t+1)} = \displaystyle \frac{(\rho_k^{(t+1)}\by-\X\f_k^{(t)})^T\mathbf{m}_k^{(t)}}{n_k^{(t)}},
\label{chi}
\end{equation}
\begin{equation}
\f_k^{(t+1)} = \underset{\f_k}{\argmin}
\Bigg\{\displaystyle\frac{1}{2} 
\Vert \M_k^{{1}/{2}(t)}(\rho_k^{(t+1)}\by-\chi_k^{(t+1)}\mathbf{1}_{n}-\X\f_k)\Vert_2^2+\lambda_{k}^{(t+1)}\Vert\f_k\Vert_1\Bigg\}.
\label{phi}
\end{equation}
Finally, the EM updates of $\thitaY$ under the NJ prior are the following
\begin{equation}
\sigma_k^{2(t+1)}=
\displaystyle \frac{\big(\by-\ak^{(t)}\mathbf{1}_{n}-\X\bbk^{(t)}\big)^{T}\M_k^{(t)}\big(\by-\ak^{(t)}\mathbf{1}_{n}-\X\bbk^{(t)}\big)}{n_k^{(t)}+2},
\label{sigma}
\end{equation}
\begin{equation}
\ak^{(t+1)} = 
\displaystyle \frac{(\by-\X\bbk^{(t)})^T\mathbf{m}_k^{(t)}}{n_k^{(t)}},
\label{alpha}
\end{equation}
\begin{equation}
\bbk^{(t+1)}  		
=\Big(\X^{T}\M_k^{(t)}\X+\sigma_k^{2(t+1)}\V_k^{(t)}\Big)^{-1}\X^{T}\M_k^{(t)}\Big(\by-\ak^{(t+1)}\mathbf{1}_{n}\Big).
\label{beta}
\end{equation}
As remarked previously, in practice we work with $\U=\V_k^{-1(t)}$
Specifically, we have two available options for $\bbk$; the first is suited for the $n>p$ case and is given by
\begin{equation}
\bbk^{(t+1)} =
\Usqrt\Big(\sigma_k^{2(t+1)}\Ip+\Usqrt\X^{T}\M_k^{(t)}\X\Usqrt\Big)^{-1}\Usqrt\X^{T}\M_k^{(t)}\Big(\by-\ak^{(t+1)}\mathbf{1}_{n}\Big), 
\end{equation}
while the second, which is faster to compute when $n<p$, is given by
\begin{equation}
\bbk^{(t+1)}  
=
\sigma_k^{-2(t+1)}\U\Big[\Ip-\X^T\Big(\sigma_k^{2(t+1)}\M_k^{-1(t)}+\X\U\XT\Big)^{-1}\X\U\Big]
\X^{T}\M_k^{(t)}\Big(\by-\ak^{(t+1)}\mathbf{1}_{n}\Big)
\end{equation}
Additional details on practical implementation appear in Appendix D.

\subsection{Convergence guarantees}

\subsubsection{Preliminaries}
The proposed EM algorithm is an expectation/conditional-maximization (ECM) as introduced by \cite{Meng_Rubind1993}. Let us recall some elements of their formalism. We call $\boldsymbol\xi \in \Xi$ the variable optimised in the M step. The corresponding optimised function is $Q(\boldsymbol\xi | \boldsymbol\xi^{(t)})$, where $\boldsymbol\xi^{(t)}$ is the value of the parameter after $t$ ECM steps. Then, the exact M step is defined as 
\begin{equation} \label{eq:exact_M_step}
\boldsymbol\xi^{(t+1)} := \underset{\boldsymbol\xi \in \Xi}{\argmax}\, Q(\boldsymbol\xi | \boldsymbol\xi^{(t)}) \, .
\end{equation}
When the optimisation in \eqref{eq:exact_M_step} is inconvenient, \cite{Meng_Rubind1993} proposed to replace it by $S\in \mathbb{N}^*$ successive block-wise updates (``conditional maximization; CM). Given $S$ constraint functions $\brace{g_s(\boldsymbol\xi)}_{s=1}^S$, the CM step is decomposed into the $S$ intermediary steps:
\begin{equation} \label{eq:CM_step}
\boldsymbol\xi^{(t+s/S)} := \underset{\boldsymbol\xi \in \Xi; g_s(\boldsymbol\xi)=g_s(\boldsymbol\xi^{(t+(s-1)/S)})}{\argmax}\, Q(\boldsymbol\xi | \boldsymbol\xi^{(t)}) \, ,
\end{equation}
for $s = 1, ..., S$.
In their theorems 2 and 3, \cite{Meng_Rubind1993} provide conditions under which all limit points of any ECM sequence are critical points of the observed likelihood. We propose a reformulation of their theorem 3, where we list explicitly all the required conditions.  
\begin{theorem}[Theorem 3 of \cite{Meng_Rubind1993}]\label{thm:convergence_ECM}
	With $r \in \mathbb{N}^*$, let $\Xi$ be a subset of the Euclidean space $\R^r$. Let $\big\{\boldsymbol\xi^{(t)}\big\}_{t\in \mathbb{N}} \in \Xi^{\mathbb{N}}$ be an ECM sequence that has an observed log-likelihood called $L(\boldsymbol\xi)$ as its objective function. The initial term $\boldsymbol\xi^{\scaleto{(0)}{6pt}}$ is such that $L(\boldsymbol\xi^{\scaleto{(0)}{6pt}})>-\infty$. Let $Q(\boldsymbol\xi | \boldsymbol\xi^{(t)})$ be the corresponding expected complete likelihood that is conditionally maximised at each CM step, with constraints functions $\brace{g_s(\boldsymbol\xi)}_{s=1}^S$. Finally, call $\Xi^{\circ}$ the interior of $\Xi$, and assume that:
	\begin{itemize}
		\item {Each conditional maximisation in the CM step \eqref{eq:CM_step} has a unique optimum}
		\item $\forall s, g_s(\boldsymbol\xi)$ is differentiable and the gradient $\nabla g_s(\boldsymbol\xi)\in \R^{r \times d_s}$ is of full rank on $\Xi^{\circ}$
		\item $\bigcap_{s=1}^S \brace{\nabla g_s(\theta) u  | u \in \R^{d_s}} = \{0\}$
		\item The condition (6)-(10) of \cite{wu1983convergence}:
		\begin{itemize}
			\item[(6)] $\Xi_{\boldsymbol\xi^{(0)}} := \brace{\boldsymbol\xi \in \Xi | L(\boldsymbol\xi) \geq L(\boldsymbol\xi^{(0)})}$ is compact for any $L(\boldsymbol\xi^{(0)})>-\infty$
			\item[(7)] $L$ is continuous on $\Xi$ and differentiable on $\Xi^{\circ}$
			\item[(9)] $\Xi_{\boldsymbol\xi^{(0)}} \subseteq \Xi^{\circ}$
			\item[(10)] $Q(\boldsymbol\xi_1 | \boldsymbol\xi_2)$ is continuous in both $\boldsymbol\xi_1$ and $\boldsymbol\xi_2$.
		\end{itemize}
	\end{itemize}
	Then all limit points of $\big\{\boldsymbol\xi^{(t)}\big\}_{t\in \mathbb{N}}$ are stationary points of the objective $L(\boldsymbol\xi).$
	
\end{theorem}
Note that condition (8) of \cite{wu1983convergence}:
\begin{itemize}
	\item[(8)] The sequence $\big\{L(\boldsymbol\xi^{(t)})\big\}_t$ {is upper bounded} for any $\boldsymbol\xi^{(0)} \in \Xi$
\end{itemize}
is verified as a direct consequence of (6) and (7) and is actually not an additional condition.

\subsubsection{Main results} \label{sect:main_convergence_result}
Here, we apply the convergence Theorem \ref{thm:convergence_ECM} to the proposed ECM algorithm for the RJM model. First we show that without modification, our ECM verifies almost all the hypotheses of Theorem \ref{thm:convergence_ECM}. In particular the ones specific to the ECM procedure, as laid out by \cite{Meng_Rubind1993}. Then, we provide conditions on the ECM penalization under which the remaining, more restrictive, regularity hypotheses in \cite{wu1983convergence} are also verified. 

In our case, the optimization variable is $\boldsymbol\xi := (\m, \Om, \f, \boldsymbol\chi, \boldsymbol\rho, \boldsymbol\lambda, \boldsymbol\tau)\in \Xi$. 
Where the closure of the parameter set $\Xi$ is $\overline{\Xi} = \R^{Kp} \times S_p(\R)^{+} \times \R^{Kp} \times \R^{K} \times {\R_+}^K \times {\R_+}^K \times S_K \subset \R^{K(p^2+2p+3)}$, with $S_p(\R)^{+}$ the cone of positive semi-definite matrices of size $p$ and $S_K := \brace{\boldsymbol\tau \in [0, 1]^K | \sum_k \tau_k=1}$. Its interior is $\Xi^{\circ} = \R^{Kp} \times S_p(\R)^{++} \times \R^{Kp} \times \R^{K} \times {\R_+^{*}}^K \times {\R_+^{*}}^K \times S_K^{\circ}$, with $S_p(\R)^{++}$ the open cone of positive definite matrices of size $p$ and $S_K^{\circ} := \brace{\boldsymbol\tau \in ]0, 1[^K | \sum_k \tau_k=1}$.
The ECM sequence takes its values in $\Xi$. With the proper priors, the parameters $\Om, \boldsymbol\rho, \boldsymbol\lambda, \text{ and } \boldsymbol\tau$ cannot take values on the border of their respective sets during an ECM sequence. In such a scenario, the ECM sequence lives in $\Xi^{\circ}$ and we can simply consider that $\Xi=\Xi^{\circ}$, which helps with several of the hypotheses. 
To ensure this property, it is sufficient to set the regularization such that the objective $L(\boldsymbol\xi)$ is infinite everywhere on the border.
This objective is the penalized observed log-likelihood function:
\begin{equation} \label{eq:RJM_observed_likelihood}
\begin{split}
L(\boldsymbol\xi) &= \sum_{i=1}^{n}\log \parent{\sum_{k=1}^{K} p(y_i|\thitaY,\xxi,z_i=k) p(\xxi|\thitaX,z_i=k)\tau_k} - \mathrm{pen}(\boldsymbol\xi)  \\
&= \sum_{i=1}^{n} \log \sum_{k=1}^{K} \exp\Bigg( -\frac{1}{2} \bigg( (y_i \rho_k - \chi_k - \xii^T \f_k)^2 -2 \log \rho_k\\
&\hspace{4.2cm}+ \norm{\xii-\m_k}_{\Om_k}^2 - \log \det{\Om_k} \\
&\hspace{4.2cm} -2\log\tau_k + (p+1) \log 2\pi + \frac{2}{n} \mathrm{pen}(\boldsymbol\xi) \bigg)\Bigg), 
\end{split}
\end{equation}
where $\norm{\xii-\m_k}_{\Om_k}^2 := (\xii-\m_k)^T\Om_k(\xii-\m_k)$. The form $\mathrm{pen}(\boldsymbol\xi)$ for the penalty term is a generalization of the separable penalty $\sum_{k}\mathrm{pen}(\thita^{*{\scaleto{X}{4pt}}})+\sum_{k}\mathrm{pen}(\thita^{*{\scaleto{Y}{4pt}}})$ proposed in Eq.~\eqref{pen_est}.
With the posterior weights $m^{(t)}_{ki}=\widehat{\Pr}(z_i=k|y_i,\xxi,\xi^{(t)})$ as defined in the E-step \eqref{member}, we can define the function $Q(\boldsymbol\xi | \boldsymbol\xi^{(t)})$ to maximise in the CM step:
\begin{equation} \label{eq:RJM_expected_complete_likelihood}
\begin{split}
Q(\boldsymbol\xi | \boldsymbol\xi^{(t)}) &= \sum_{i=1}^{n}  \sum_{k=1}^{K}  -\frac{1}{2} m^{(t)}_{ki} \bigg( (y_i \rho_k - \chi_k - \xii^T \f_k)^2 -2 \log \rho_k\\
&\hspace{3.3cm}+ \norm{\xii-\m_k}_{\Om_k}^2 - \log \det{\Om_k} \\
&\hspace{3.3cm} -2\log\tau_k + (p+1) \log 2\pi + \frac{2}{n} \mathrm{pen}(\boldsymbol\xi) \bigg).
\end{split}
\end{equation}
The conditional maximization of this function is carried out in Eq.~\eqref{tau} to \eqref{phi}, for one specific version of the penalty $\mathrm{pen}(\boldsymbol\xi)$. As required by Theorem \ref{thm:convergence_ECM}, each of these optmizations is uniquely defined. This property is penalty dependent. Hence, in general it is required to use penalties/priors on the parameters that lead to uni-modal posterior distributions. 

On the other hand, the general structure of the conditional updates is independent of the penalty. Indeed, we propose a block-type update where each block is updated conditionally to all other being fixed. The order of the updates is: $\boldsymbol \tau \to \m \to \Om \to \boldsymbol\lambda \to \boldsymbol\rho \to \boldsymbol\chi \to \f$. This correspond to constraint functions of the form:
\begin{equation}\label{eq:ECM_constraints}
\begin{split}
g_1(\boldsymbol\xi) &= \boldsymbol\xi \setminus \boldsymbol\tau := (\m, \Om, \boldsymbol\lambda, \boldsymbol\rho, \boldsymbol\chi, \f)\, ,\\
g_2(\boldsymbol\xi) &= \boldsymbol\xi \setminus \m := (\boldsymbol \tau, \Om, \boldsymbol\lambda, \boldsymbol\rho, \boldsymbol\chi, \f)\, ,\\
g_3(\boldsymbol\xi) &= \boldsymbol\xi \setminus \Om :=(\boldsymbol \tau, \m, \boldsymbol\lambda, \boldsymbol\rho, \boldsymbol\chi, \f)\, ,\\
g_4(\boldsymbol\xi) &= \boldsymbol\xi \setminus \boldsymbol\lambda :=(\boldsymbol \tau, \m, \Om, \boldsymbol\rho, \boldsymbol\chi, \f)\, ,\\
g_5(\boldsymbol\xi) &= \boldsymbol\xi \setminus \boldsymbol\rho :=(\boldsymbol \tau, \m, \Om, \boldsymbol\lambda, \boldsymbol\chi, \f)\, ,\\
g_6(\boldsymbol\xi) &= \boldsymbol\xi \setminus \boldsymbol\chi :=(\boldsymbol \tau, \m, \Om, \boldsymbol\lambda, \boldsymbol\rho, \f)\, ,\\
g_7(\boldsymbol\xi) &= \boldsymbol\xi \setminus \f :=(\boldsymbol \tau, \m, \Om, \boldsymbol\lambda, \boldsymbol\rho, \boldsymbol\chi) \, .
\end{split}
\end{equation}
When the joint optimization in several consecutive blocks is possible, usually because they are separate in the objective, this approach can be simplified by ``fusing" the corresponding blocks. For instance, in Eq~\eqref{tau} to \eqref{rho}, we perform a joint optimization in $\boldsymbol \tau, \m, \Om, \boldsymbol\lambda, \boldsymbol\rho$ under the constraint that $\f, \boldsymbol \chi$ is fixed. Then, in Eq~\eqref{chi}, an optimization on $\boldsymbol \chi$ with $\boldsymbol\lambda, \boldsymbol\rho, \f$ fixed. Finally, in Eq~\eqref{phi}, an optimization on $\f$ with $\boldsymbol\lambda, \boldsymbol\rho, \boldsymbol \chi$ fixed. Note that in this case, the optimization in $\boldsymbol \tau, \m, \Om$ is a true M step and is independent on the other parameters. Hence, once this block has been updated in the first step, constraining it to remain fixed in the subsequent steps is unnecessary. 

The functions $g_s(\boldsymbol\xi)$ defined in \eqref{eq:ECM_constraints} are obviously differentiable on $\Xi$ with gradients of the form: 
\begin{equation*}
\nabla g_s(\boldsymbol\xi) = \begin{bmatrix}
0_{d_s \times d_1}  & \hdots  & 0_{d_s \times d_{s-1}}  & I_{d_s} & 0_{d_s \times d_{s+1}}  & \hdots  & 0_{d_s \times d_S} 
\end{bmatrix}^T \in \R^{r \times d_s} \, ,
\end{equation*}
which are of rank $d_s$ (full rank), with $r:=\sum_s d_s = K(p^2+2p+3)$. We can also see that there is no ``overlap" between their non-zero components, which results in the desired property that
$\bigcap_{s=1}^S \brace{\nabla g_s(\theta) u  | u \in \R^{d_s}} = \{0\}$.
As a consequence, as long as the penalty is chosen such that the posterior distribution in each parameter is unimodal, the algorithm verifies the three ``ECM-specific" conditions for convergence introduced by \cite{Meng_Rubind1993}.

Among the basic conditions identified by \cite{wu1983convergence}, some are also systematically verified by our algorithm with little assumption on the penalty; namely:
\begin{itemize}
	\item[(7)] The model part of $L(\boldsymbol\xi)$ is always continuous and differentiable in $\Xi = \Xi^{\circ}$. Hence, this property is guaranteed for $L(\boldsymbol\xi)$ as long as the penalty term is also continuous and differentiable.
	\item[(9)] $\Xi_{\boldsymbol\xi^{(0)}} := \brace{\boldsymbol\xi \in \Xi | L(\boldsymbol\xi) \geq L(\boldsymbol\xi^{(0)})} \subseteq \Xi = \Xi^{\circ}$.
	\item[(10)] $Q(\boldsymbol\xi_1 | \boldsymbol\xi_2)$ is continuous in $\xi_1$ for the same reason that $L(\boldsymbol\xi)$ is continuous in $\xi$. The dependency of $Q(\boldsymbol\xi_1 | \boldsymbol\xi_2)$ in $\boldsymbol\xi_2$ is entirely through the terms $p(z_i=k|y_i,\xxi,\boldsymbol\xi_2) = {p(y_i,\xxi, z_i=k|\boldsymbol\xi_2)}\mathbin{/}{\sum_l p(y_i,\xxi, z_i=l|\boldsymbol\xi_2)}  $ which are continuous in $\boldsymbol\xi_2$ for the same reason that the likelihood is.
\end{itemize}

The final missing hypothesis is (6), the compacity of the level lines of the likelihood function. This property is much more restrictive and requires specific hypotheses on the regularization. The following theorem synthesizes every observation made so far and provides sufficient conditions to verify the final hypothesis.
\begin{theorem}[Convergence of the ECM algorithm for RJMs] \label{thm:convergence_ECM_RJM}
	Consider an ECM sequence $\big\{\boldsymbol\xi^{(t)}\big\}_{t\in \mathbb{N}} \in \Xi^{\mathbb{N}}$ with objective function the observed log-likelihood $L(\boldsymbol\xi)$ of the RJM model \eqref{eq:RJM_observed_likelihood}. The initial term $\boldsymbol\xi^{\scaleto{(0)}{6pt}}$ is such that $L(\boldsymbol\xi^{\scaleto{(0)}{6pt}})>-\infty$ and the conditional maximization step of the expected complete likelihood $ Q(\boldsymbol\xi | \boldsymbol\xi^{(t)})$ in Eq.~\eqref{eq:RJM_expected_complete_likelihood} is conducted using the block-wise scheme defined by the constraint function $g_s(\boldsymbol\xi)$ in Eq.~\eqref{eq:ECM_constraints}. Assume that the regularization term $\mathrm{pen}(\boldsymbol\xi)$ is continuous, differentiable and such that each of the block maximizations is unique (uni-modal posterior). Assume additionally that there exists a positive constant $\delta > 0$ such that 
	\begin{equation}\label{eq:sufficient_penalty_lower_bound}
	\mathrm{pen}(\boldsymbol{\xi})\geq \delta \sum_{k=1}^{K} \parent{\log \tau_k^{-1} + \norm{\mu_k} + \norm{\Om_k} + \log \det{\Om_k^{-1}}  +  f_{\lambda}(\lambda_k) + \rho_k + \log\rho_k^{-1} +   \det{\chi_k} + \norm{\f_k}}, 
	\end{equation}
	where $f_{\lambda}$ is a lower bounded function on $\R_+^*$ such that $f_{\lambda}(x) \underset{x \to 0}{\longrightarrow} +\infty$ and $f_{\lambda}(x) \underset{x \to +\infty}{\longrightarrow} +\infty$. \\
	Then, Theorem \ref{thm:convergence_ECM} applies and all limit points of $\big\{\boldsymbol\xi^{(t)}\big\}_{t\in \mathbb{N}}$ are stationary points of the objective $L(\boldsymbol\xi).$
\end{theorem}

\begin{remark}
	~
	\begin{itemize}
		\item The norms $\norm{.}$ on each parameters in Eq.~\eqref{eq:sufficient_penalty_lower_bound} (in Appendix E) are unspecified since all norms are equivalent in finite dimension.
		\item For $f_{\lambda}$, the lower bound on the penalty in $\lambda_k$, a function such as $f_{\lambda}(x) = x - \log x$ is suitable.
	\end{itemize}
\end{remark}

{\bf Sketch of proof:} 
The full details of the proof can be found in Appendix E, providing here a brief proof sketch. We prove that under the assumptions of Theorem \ref{thm:convergence_ECM_RJM}, all the hypotheses of Theorem \ref{thm:convergence_ECM} are met, which yields the desired convergence result.

As previously discussed, all the hypotheses of Theorem \ref{thm:convergence_ECM}, save for hypothesis (6), of \cite{wu1983convergence} are organically verified within the RJM model. Provided that $L(\boldsymbol\xi)$ is infinite on the border of $\Xi$, we can safely take $\Xi = \Xi^{\circ}$, and then the few required assumptions on the penalty are verified: it is continuous, differentiable and such that each of the block maximizations is unique.
Hence, all the efforts of the proof are spent on proving that $L(\boldsymbol\xi)$ is infinite on the border of $\Xi$ and that hypothesis (6) is verified. This is done all at once; thanks to the control \eqref{eq:sufficient_penalty_lower_bound}, we are able to define an increasing family $\Xi_m$ of compacts of $\Xi^{\circ}$ such that the log-likelihood $L(\boldsymbol\xi)$ on any point of $\overline{\Xi} \setminus \Xi_m$ is as low as desired with a well chosen compact $\Xi_m$. With this result, we have that (i) $L(\boldsymbol\xi)$ is $-\infty$ outside of $\Xi^{\circ}$
and (ii)  $\Xi_{\scaleto{\boldsymbol\xi}{6pt}^{(0)}} := \big\{\boldsymbol\xi \in \Xi | L(\boldsymbol\xi) \geq L(\boldsymbol\xi^{\scaleto{(0)}{6pt}})\big\}$ is compact for any $L(\boldsymbol\xi^{\scaleto{(0)}{6pt}})>-\infty$ (hypothesis (6)), allowing us to conclude.

\section{Prediction and cluster selection}
\label{prediction}           

An interesting feature of RJMs
relates to prediction. Specifically, a new observation 
$\xstar {\in} \mathbb{R}^p$ 
can be allocated 
to a cluster via the quantities
$
\hat{\pi}_k^*~\propto~\hat{\tau}_k\varphi_p(\xstar|\hat{\boldsymbol{\mu}}_k,\hat{\boldsymbol{\Sigma}}_k),
$
where $\hat{\tau}_k$, $\hat{\boldsymbol{\mu}}_k$ and $\hat{\boldsymbol{\Sigma}}_k$ are EM estimates for $k=1,\dots,K$. A simple prediction of  $y^*$ then follows via
$
\hat{y}^*{=}\hat{\alpha}_{\tilde{k}}+{\xstar}^T\hat{\boldsymbol{\beta}}_{\tilde{k}}$, where $\tilde{k} {=} \argmax_k \hat{\pi}_k
$
and $\hat{\alpha}_{\tilde{k}}$ and $\hat{\boldsymbol{\beta}}_{\tilde{k}}$ are the corresponding EM estimates.

Although not our primary focus in this paper, it is interesting to 
briefly consider  the idea of setting $K$ based on predictive loss.
In more detail,
let $\mathcal{G}$ denote the set of the number of clusters under consideration, so that the cluster indicator under model $g\in\mathcal{G}$ is $k_g=1,\dots,K_g$. 
Under each model $g$ we can obtain cluster allocations for a subset of held-out test data $\by^*$ and $\X^*$. Denote by $\by^*_{k_g}$ the $n^*_{k_g}\times 1$ test-response vector and by $\X^*_{k_g}$ the $n^*_{k_g}\times p$ test-feature matrix assigned to group $k_g=1,\dots,K_g$ conditional on $g$. Then, the solution for the ``best'' group-wise predictive model (in an $\ell_2$ sense) is given by 
\begin{equation}
\hat{g}^{\mathrm{pred}}=\underset{g\in\mathcal{G}}{\argmin}\Bigg\{\frac{1}{K_g}\sum_{k_g=1}^{K_g}\frac{\big\Vert \by^*_{k_g} -
	\hat{\alpha}_{k_g} \mathbf{1}_{n^*_{k_g}} -
	\X^{*T}_{k_g}\hat{\boldsymbol{\beta}}_{k_g} \big\Vert_2^2}{n^*_{k_g}}\Bigg\}.
\label{ghat}
\end{equation}
This effectively sets $K$ 
to minimize predictive loss 
and connects in a way supervised and unsupervised learning,
providing a simple and quite natural way of determining the number of clusters based on a ``guided'' search that aims to optimize prediction of $\by$.

Of course, standard approaches for inferring the number of clusters, such as information criteria, can be used within the RJM framework. For instance, using BIC \citep{schwarz_78} in our case translates in selecting the number of clusters as
\begin{equation}
\hat{g}^{\mathrm{BIC}} = \underset{g\in\mathcal{G}}{\argmax} \Big\{ 2\log p(\by,\X|\boldsymbol{\theta}_g,\boldsymbol{\tau}_g)-\log(n)\nu_g \Big\},
\label{bic}
\end{equation}
where $\nu_g$ is the number of elements in $\boldsymbol{\theta}_g$ that are not set equal to zero. Similarly, for AIC \citep{akaike_74} we have  
\begin{equation}
\hat{g}^{\mathrm{AIC}} = \underset{g\in\mathcal{G}}{\argmax} \Big\{ 2\log p(\by,\X|\boldsymbol{\theta}_g,\boldsymbol{\tau}_g)-2\nu_g \Big\}.
\label{aic}
\end{equation}
\section{Empirical examples}
\label{simulations}    

\begin{figure}[h]
	\centering{}\includegraphics[scale=0.5]{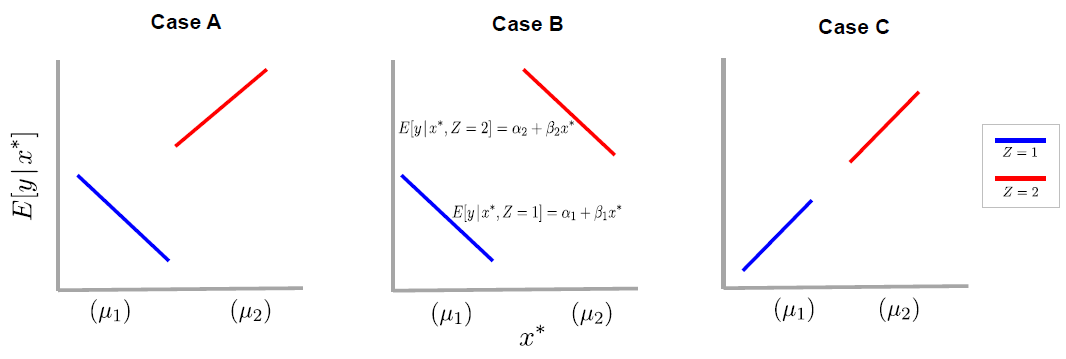}
	\caption{Some interesting cases of group structure.
		Univariate analogues of problems considered in the empirical examples (all of which are multivariate) in order to illustrate the key ideas.
		Shown 
		are two latent groups with some separation between the group-wise means of a single feature $x^*$. 
		Three specific cases, that differ with respect to the regression model linking $y$ and $x^*$, are shown: equal-intercept/unequal-slope (Case A); unequal-intercept/equal-slope (Case B);  and equal intercept and slope (Case C).
	}
	\label{illustration}
\end{figure}

In this Section we present results from simulation experiments, starting with a small-scale 
simulation in Section \ref{sim1} which allows us to evaluate and visualize easily the various learning aspects of RJMs.  
In Section \ref{sim2} we use data from The Cancer Genome Atlas in  semi-synthetic examples of much 
larger scale, providing 
detailed comparisons with baseline and various oracle-type approaches.
All simulations are based on data-generating mechanisms which are multivariate generalizations of three elementary problems which are depicted in Figure \ref{illustration}, the purpose of which is to facilitate understanding of more complex multivariate problems as the ones in Section \ref{sim2} via illustration of simpler univariate analogues. 
Finally, in Section \ref{sim4} we show results using fully empirical data.

\subsection{Small-scale simulations}
\label{sim1}

\noindent
\textbf{Set-up.} We consider two groups ($K=2$) with total $n=100$ and balanced groups, i.e. $n_k=50$ for $k = 1, 2$. The number of predictors is $p=10$, where in each group only the first predictor ($\mathbf{x}_{k1}$) has a non-zero coefficient, i.e. only $\beta_{k1}\ne0$ for $k=1,2$. The covariates are generated as
$\X_k\sim\N_p(\mm_k,\Sig_k)$, with $\mm_1=(0,\dots,0)^T$ and $\mm_2=(1,\dots,1)^T$ being of dimensionality $p\times 1$. 
For the covariances we consider two scenarios: an \textit{uncorrelated-scenario} with diagonal covariances of the form $\Sig_1=\Sig_2=\Ip$ and a \textit{correlated-scenario} with non-diagonal covariances, where  each variable $\mathbf{x}_{kj}$, for $j=2,\dots,p$, is again Gaussian noise, but the signal variable $\mathbf{x}_{k1}$ is generated as 
$
\mathbf{x}_{k1} \sim \N_{n_k}\big(1.5\mathbf{x}_{k3}+0.5\mathbf{x}_{k5}-0.7\mathbf{x}_{k7},~ 0.5 \mathbf{I}_{n_k}\big).
$   
The response of each group is generated as
$
\by_k\sim \N_{n_k}\big(\mathbf{1}_{n_k}\alpha_k+\mathbf{x}_{k1}\beta_{k1},~ \sigma_k^2\mathbf{I}_{n_k}\big). 
$
Specification of the slopes and intercepts 
is based on
the three cases 
of Figure \ref{illustration}; see Table \ref{sim1_param} (Appendix F).
Finally, the error variance $\sigma_k^2$ of each group is set 
to fix signal strength in a label-oracle sense, namely 
so that the correlation between test data (for test group sample sizes of 250) and predictions from a lasso model 
is approximately 0.8 when group labels are known. The results that follow are from 50 repetitions of each simulation. 
We focus on regression signal detection, estimation of  coefficients and group assignment performance.     

\medskip

\begin{figure}[h]
	\centering{}\includegraphics[scale=0.35]{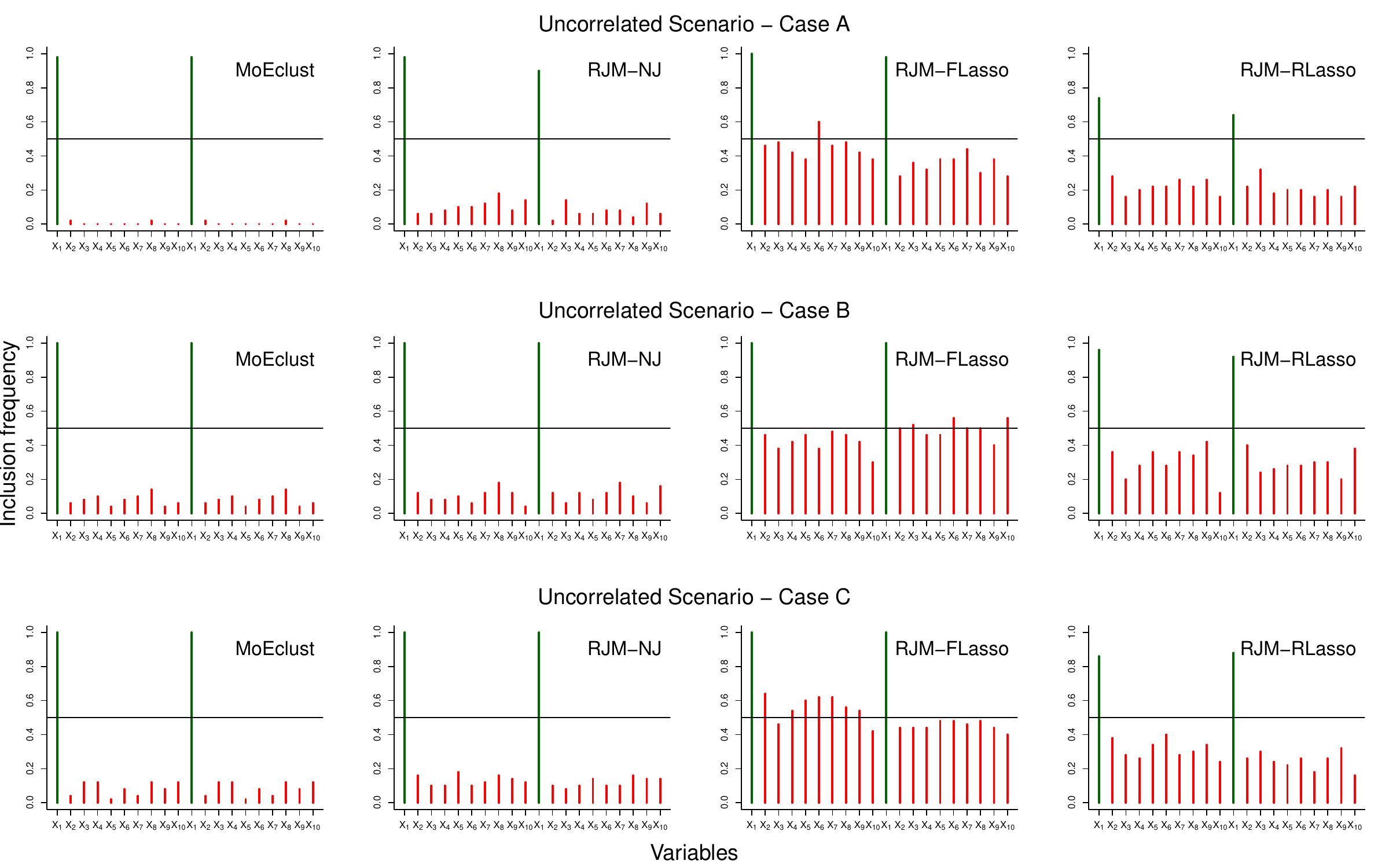}
	\caption{First simulation, uncorrelated scenario. Variable inclusion frequencies (under 50 repetitions) for signal variables (in green) and non-signal variables (in red) for regression cases A, B and C. Horizontal black lines correspond to a frequency of 0.5.}
	\label{inc.diag}
\end{figure}

\noindent \textbf{Signal detection and estimation.} The variable inclusion frequencies 
over 50 repetitions of the simulations
for the uncorrelated scenario are presented in Figure \ref{inc.diag}. RJM-NJ performs better overall as it detects influential effects almost all of the times, while the inclusion rates of non-influential effects are much lower than 50\%. RJM-FLasso is effective in detecting the signals but produces much denser solutions.
RJM-RLasso solutions 
are sparser in comparison to FLasso; we note, however, that RLasso tends to over-shrink the coefficients of the influential predictors as well. 
The inclusion frequencies for the correlated scenario are similar (Appendix F, Figure \ref{inc.corr}). 
Violin plots of slope estimates 
are presented in Appendix F  (Figure \ref{slope.diag}, uncorrelated scenario; Figure \ref{slope.corr}, correlated scenario), while 
the corresponding plots for intercepts are presented in Figures \ref{intercept.diag} and \ref{intercept.corr}. The NJ estimates are overall more accurate. In all comparisons we include results from mixtures of experts obtained from 
R package \texttt{MoEClust} \citep{murphy_murphy2020}, which performs simultaneous selection for experts, gates and covariance structures using forward search model selection based on BIC. In terms of variable selection MoE performs exceptionally well under case A and yields similar results to RJM-NJ under cases B and C (see Figures \ref{inc.diag} and \ref{inc.corr}). As for estimation, MoE leads to overall accurate slope and intercept estimates in case A, however, the corresponding estimates in cases B and C have a higher variance in comparison to RJM estimates (see Figures \ref{slope.diag} to \ref{intercept.corr}, Appendix F).  

\begin{figure}
	\centering{}\includegraphics[scale=0.5]{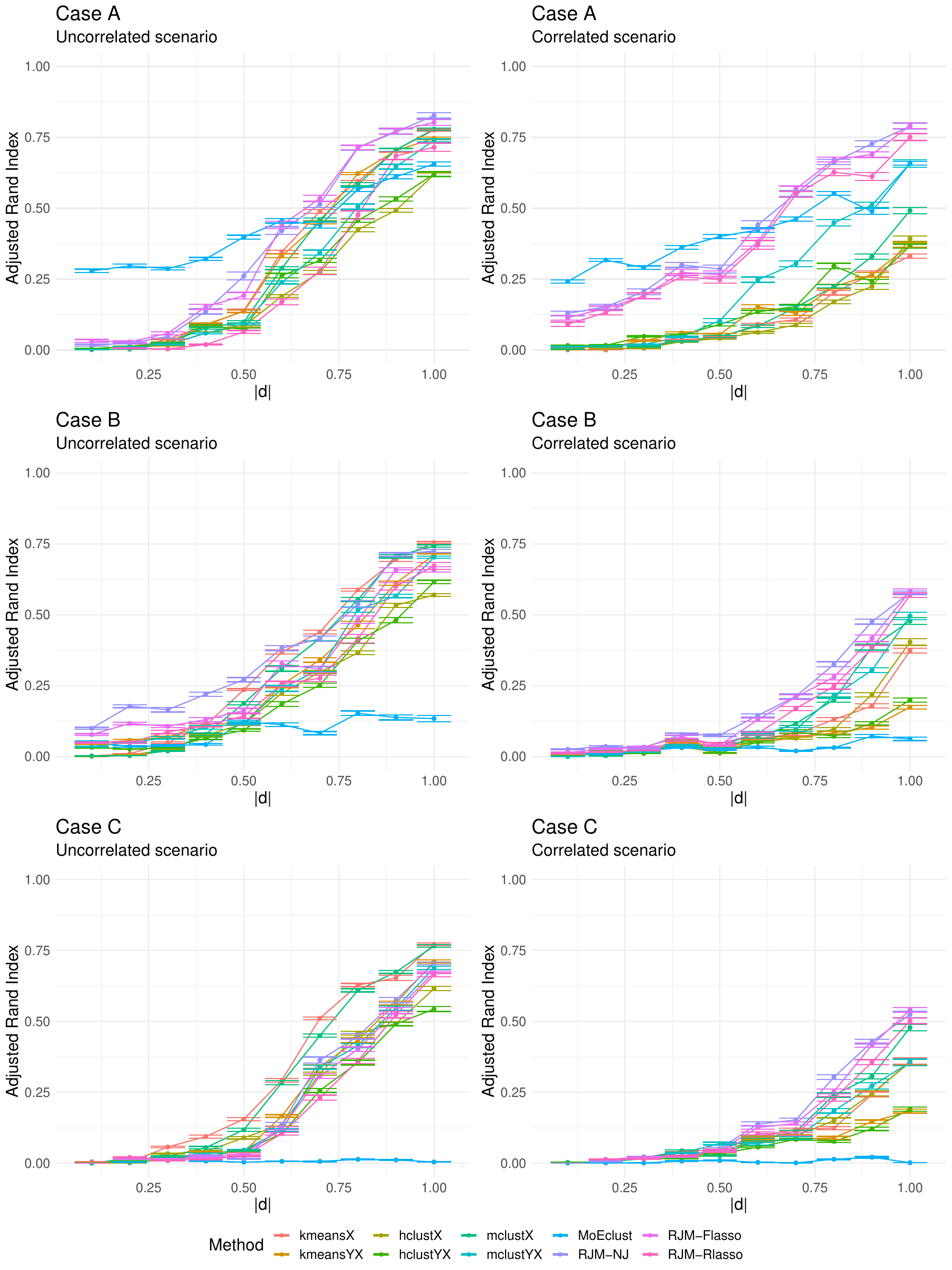}
	\caption{First simulation, cases A (top), B (middle) and C (bottom). One standard-error plots of adjusted Rand index averages (from 20 repetitions) vs  absolute distance ($|d|$) of the group-wise covariate means under the uncorrelated scenario (left) and correlated scenario (right). }
	\label{rand_compA}
\end{figure}

\medskip

\noindent \textbf{Latent group assignment.}  A natural question is whether including the regression part of the model within a unified framework provides any 
gains with respect to simply clustering the $X$ matrix. In practice, between-group differences in means may be subtle. Hence, we consider in particular performance as the magnitude of the mean difference varies; 
i.e., we set $\mm_1=\mathbf{0}_p$ and $\mm_2=\mathbf{0}_p+d\mathbf{1}_p$ with  $d=h\cdot U$, where $h$ defines a grid ranging from 0.1 to 1 with a step size of 0.05 and $U {\in} \{ -1, +1 \}$ is a uniformly random sign. Hence, $|d|$ is a measure of the strength of the mean  signal. 
We compare to $k$-means, hierarchical clustering, GMMs and MoEs as implemented in R using the default options of \texttt{kmeans}, \texttt{hclust}, \texttt{mclust} \citep{mclust} and \texttt{MoEClust} respectively; for the latter two using the BIC-optimal model. In addition, for the clustering approaches we use as input: (i) only $\X$ and (ii) $\X$ together with $\by$ stacked in one data matrix. 
For these simulations we use 20 repetitions.

One standard-error plots of
adjusted Rand index averages 
as functions of $|d|$ are shown in Figure \ref{rand_compA}. As seen, in case A, RJMs  
generally outperform all methods except of MoE; the latter performs better for lower values of $|d|$, while RJMs perform better for higher values.
In cases B and C 
(where RJM is over-parameterized) our methods remain competitive in the uncorrelated scenario, while lead to better overall results in the correlated scenario. On the other hand, MoE is not effective under cases B and C, which as argued in Section \ref{intro} (recall Figure \ref{fig_intro}) is to be expected when there are no differences in regression coefficients.

\subsection{Semi-synthetic simulations based on real cancer data}
\label{sim2}

The simulations presented below are based on data from the The Cancer Genome Atlas (TCGA, \url{https://cancergenome.nih.gov}).
The rationale is to anchor the simulation in real covariance structures. Specifically, we use data previously used in \cite{Bernd2019}, consisting of gene expression values from four cancer types; breast (BRCA), kidney renal clear cell (KIRC), lung adenocarcinoma (LUAD) and thyroid (THCA). 
Our 
strategy is to treat the cancer type as hidden: this allows us to test our approaches in the context of differential covariance structure as seen in a real group-structured problem whilst having access to true gold-standard labels.

\begin{table}[h]
	\caption{Second simulation. Intercept values and slope-generating mechanisms for the two groups under the three cases illustrated in Figure \ref{illustration}.
		[$\mathrm{TN}(\mu,\sigma^2,l,u)$ denotes a truncated normal distribution, where $\mu \in \mathbb{R}$, $\sigma^2> 0$ and $l$, $u$ are the respective lower and upper truncation bounds, while  $\mathrm{mTN}(\mu,\sigma^2,a,b)$ denotes the specific mixture of $\mathrm{TN}(\mu,\sigma^2,-\infty,a)$ and $\mathrm{TN}(\mu,\sigma^2,b,\infty)$ with $a<b$ and mixing parameter equal to 0.5; i.e., a truncated normal with support everywhere except in ($a,b$)].}
	
	\centering
	\scalebox{0.9}{%
		\begin{tabular}{ccccc}
			\hline
			\noalign{\vspace{0.18cm}}
			\multirow{2}{*}{\textbf{Case}} & \multirow{2}{*}{\textbf{Group}} & \multirow{2}{*}{\textbf{Intercept}} & \multicolumn{2}{c}{\textbf{Slopes}}\tabularnewline
			\cline{4-5}
			\noalign{\vspace{0.14cm}} 
			&  &  & \textbf{Common locations} & \textbf{Disjoint locations}\tabularnewline[\doublerulesep]
			\hline 
			\noalign{\vspace{0.1cm}} 
			\multirow{2}{*}{A} & 1 & \multirow{2}{*}{$\alpha_1=\alpha_2=0$} & $\beta^*_1\sim\mathrm{TN}(0,\tilde{\sigma}^2,-\infty,-0.1)$ & $\beta^*_1\sim\mathrm{mTN}(0,\tilde{\sigma}^2,-0.1,0.1)$
			\tabularnewline
			& 2 &  & $\beta^*_2\sim\mathrm{TN}(0,\tilde{\sigma}^2,0.1,\infty)$ & $\beta^*_2\sim\mathrm{mTN}(0,\tilde{\sigma}^2,-0.1,0.1)$ \tabularnewline
			\hline
			\noalign{\vspace{0.1cm}}  
			\multirow{2}{*}{B} & 1 & $\alpha_1=0$ & \multirow{2}{*}{$\beta^*_1=\beta^*_2\sim\mathrm{mTN}(0,\tilde{\sigma}^2,-0.1,0.1)$} & $\beta^*_1\sim\mathrm{mTN}(0,\tilde{\sigma}^2,-0.1,0.1)$ \tabularnewline
			& 2 & $\alpha_2=1$ &  & $\beta^*_2\sim\mathrm{mTN}(0,\tilde{\sigma}^2,-0.1,0.1)$\tabularnewline
			\hline
			\noalign{\vspace{0.1cm}} 
			\multirow{2}{*}{C} & 1 & \multirow{2}{*}{$\alpha_1=\alpha_2=0$} & \multirow{2}{*}{$\beta^*_1=\beta^*_2\sim\mathrm{mTN}(0,\tilde{\sigma}^2,-0.1,0.1)$} & $\beta^*_1\sim\mathrm{mTN}(0,\tilde{\sigma}^2,-0.1,0.1)$ \tabularnewline
			& 2 &  &  &$\beta^*_2\sim\mathrm{mTN}(0,\tilde{\sigma}^2,-0.1,0.1)$ \tabularnewline
			\hline
	\end{tabular}}
	\label{sim2_param}
\end{table}
\medskip
\noindent\textbf{Set-up.} In these experiments we use covariates from two cancer types; namely, the BRCA and KIRC groups. For all simulations we use $n=250$, balanced group sample sizes, i.e. $n_k=125$ for $k=1,2$, and varying dimensionality for the features; namely, i) $p=100$ ($n>p$ problem), ii) $p=250$ ($n=p$ problem) and iii) $p=500$ ($n<p$ problem). We consider sparse problems where the percentage of non-zero  coefficients ($\bb^*$)  is $s=4\%$ 
and the 
setting in which some of the non-zero coefficients are at common locations and others are at disjoint locations across the two groups (placing half the non-zero coefficients at common locations). 
Specification of the common-location $\bb_k^*$'s will determine the three general cases depicted initially in Figure \ref{illustration}. 
To rule out very small coefficients, we draw from a  truncated normal distribution, with support excluding the interval $(-0.1,0.1)$.
Group specific intercept values and slope-generating mechanisms (based on 
Figure \ref{illustration}), are summarized in Table \ref{sim2_param}. 
Given the matrices $\X_k$, the intercepts $\ak$ and the sparse vectors $\bb_k$ the response is generated as 
$
\by_k\sim\mathrm{N}_{n_k}(\boldsymbol{m}_k,~\mathbf{I}_{n_k}\sigma^2_y),
$
where $\boldsymbol{m}_k=\mathbf{I}_{n_k}\ak+\X_k\bb_k$ for $k=1,2$ and $\sigma^2_y=1$. The scale parameter $\tilde{\sigma}^2$ in Table \ref{sim2_param} is tuned so that the overall signal-to-noise under each case is approximately equal to three; i.e. $\mathrm{Var}(\boldsymbol{m})/\sigma_y^2\approx3$ and $\boldsymbol{m}=(\boldsymbol{m}_1,\boldsymbol{m}_2)^T$.

Performance is evaluated as a function of the absolute distance $|d|$ between the group-wise feature means $\mm_1$ and $\mm_2$.
We initially normalize the features, so that $\mm_1=\mm_2=\mathbf{0}$, and consider again the case where each element of $\mm_2$ is shifted by $d=h\cdot u$, where $h\in\{0.1,0.2,\dots,0.8,0.9\}$ 
and $u$ is a random uniform sign. Each simulation is repeated 20 times using random subsamples of features from the original data. Here we present results for the $n<p$ setting ($p=500$); results for $p\in\{100,250\}$ can be found in Appendix G. 
For the regression questions addressed below our aim is to compare RJM with the 
``clustering-then-regression" approach.
Obviously, a range of regression methods could be used in the second step.
As the  simulations are sparse and linear by design, to ensure that the simple ``clustering-then-regression" approach is not disadvantaged
we use lasso in the second step. 

\medskip

\begin{figure}[h]
	\centering{}\includegraphics[height=7cm,width=0.9\textwidth]{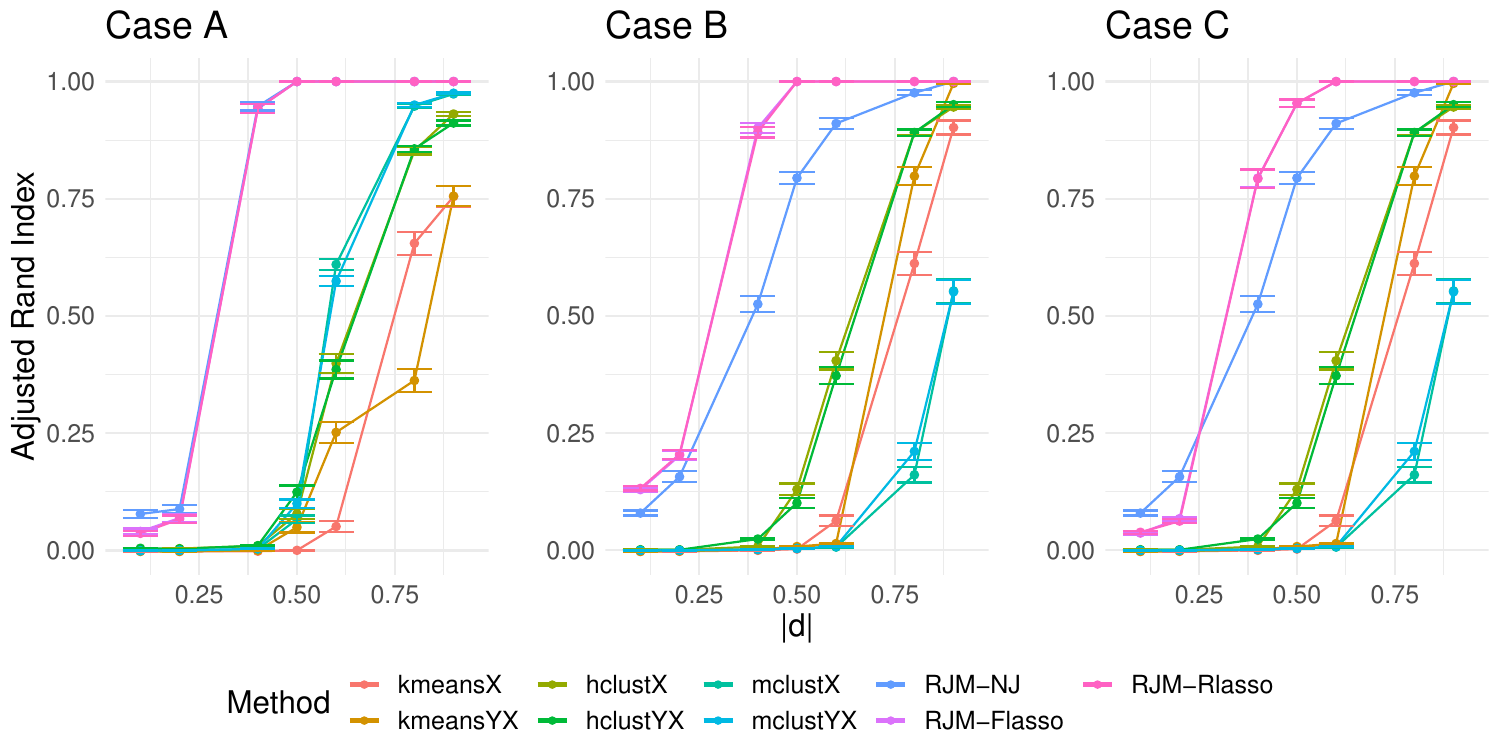}
	\caption{Second simulation, $p=500$, group assignment. Average adjusted Rand Index 
		as a function of the absolute distance ($|d|$) of the group-wise covariate means, for cases A (left), B (center) and C (right). [Error bars indicate standard errors from 20 repetitions.]}
	\label{rand500}
\end{figure}

\noindent\textbf{Group assignment.} 
We compare to the same methods considered in Section \ref{sim1}, except of \texttt{MoEClust} as the dimensionalities are too large for a forward model search for optimal selection of expert and gating functions. 
Figure \ref{rand500} presents error plots of adjusted Rand index averages.
In general, we observe a phase-transition type of behaviour as all methods improve as $|d|$ increases. However, the transition is faster with RJM which outperforms the other methods and stabilizes relatively quickly to correct assignment. For $p=500$ lasso-based RJM outperforms the NJ variant for cases B and C, however, for $p=\{100,250\}$ all RJM methods perform equally well more or less; see Figures \ref{rand100} and \ref{rand250} in Appendix G.

\medskip

\noindent\textbf{Variable selection.} For this comparison we initially set a benchmark model called the \textit{label-oracle-lasso}, under which the true group labels are assumed known and we fit separate lasso regressions 
via \verb|glmnet| \citep{glmnet}. We also consider a \textit{cluster-lasso} model which involves separate regressions 
based on estimated group labels. This approach involves an initial clustering step: we give an advantage to the cluster-lasso by using, for each dimension considered, the clustering approach that performs best
(\verb|hclust| for $p=100$ and \verb|Mclust| for $p\in\{250,500\}$). 
Naturally, the cluster-lasso will be equivalent to the oracle-lasso  when  group assignment is perfect.           

\begin{figure}[H]
	\centering{}\includegraphics[height=7cm,width=0.9\textwidth]{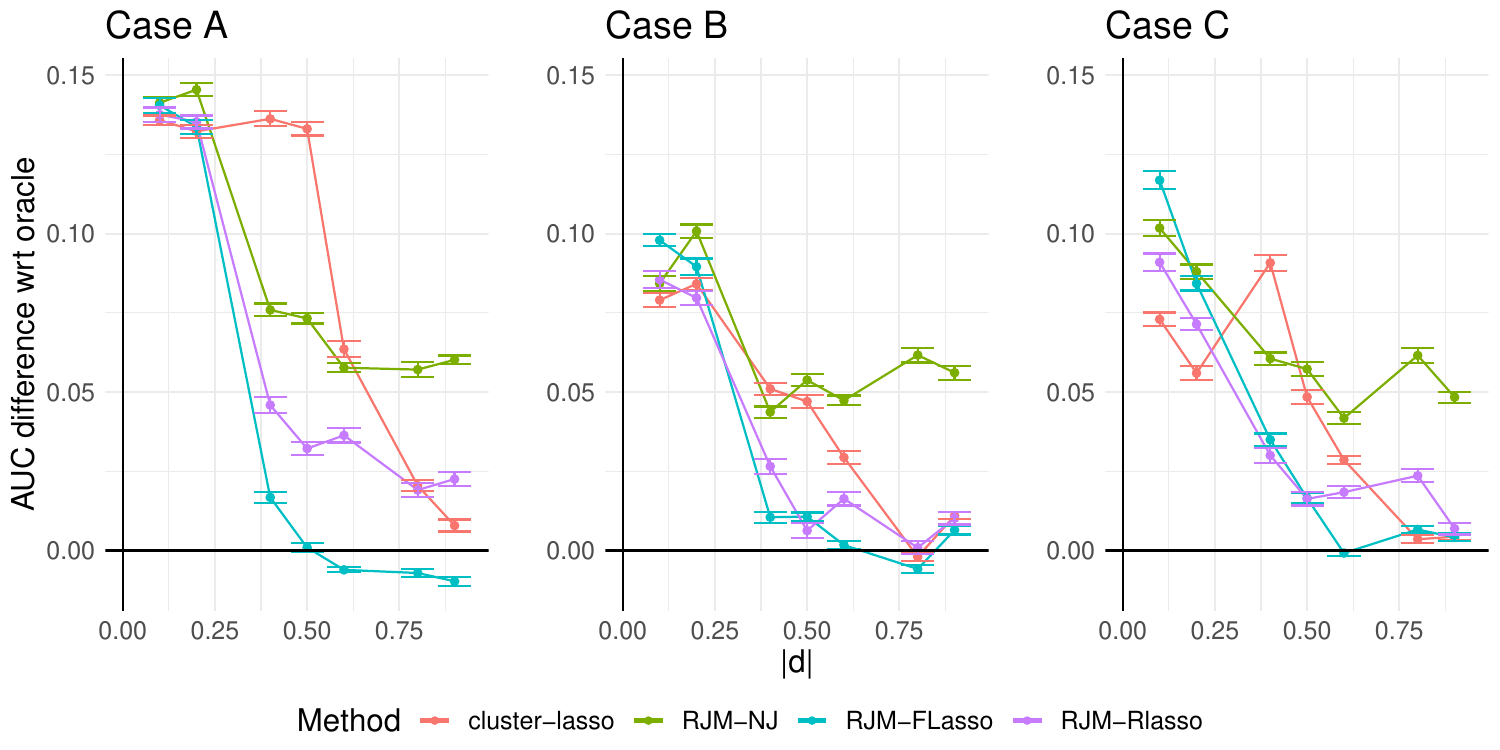}
	\caption{Second simulation, $p=500$, variable selection. AUC loss 
		from  oracle-lasso 
		as a function of  the absolute distance ($|d|$) of the group-wise covariate means, for cases A (left), B (center) and C (right). [Error bars indicate standard errors from 20 repetitions.]}
	\label{auc500}
\end{figure}

We summarize results via the area under the ROC curve (AUC) based on the ranking of the absolute values of the coefficients. In particular, we consider the difference between the AUC from oracle-lasso and AUC from competing methods (cluster-lasso and RJM approaches). 
One standard-error plots 
for $p=500$
are presented in Figure \ref{auc500}.
As expected cluster-lasso yields smaller selection loss (approaching  oracle-lasso) as the separation of group-wise means increases, but so do the RJM methods. RJM-FLasso is overall better and even seems to result in slightly improved selection in comparison to the oracle-lasso as $|d|$ increases, possibly due to the fact that RJM uses weighted estimation based on the entire sample. Importantly, RJM methods overall outperform the common cluster-lasso approach in low and/or medium magnitude regions of $|d|$. These results illustrate 
the nontrivial gains possible from a unified treatment of the various aspects of the model vs. the simple approach of clustering followed by sparse regression.  For the simulations with $p=100$ and $p=250$ see Figures \ref{auc100}  and \ref{auc250} (Appendix G), respectively.

\begin{figure}[h]
	\centering{}\includegraphics[height=8cm,width=\textwidth]{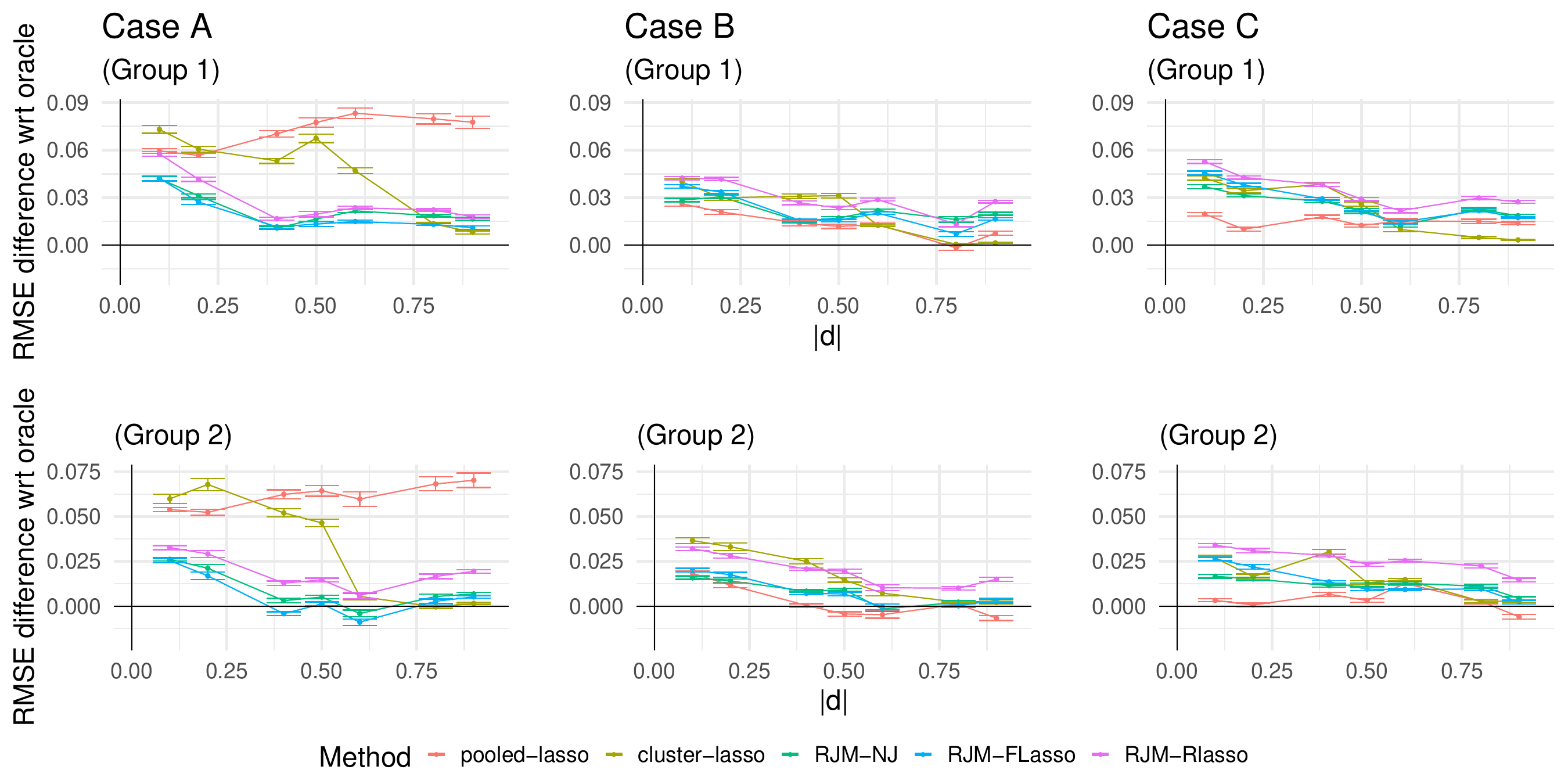}
	\caption{Second simulation, $p=500$, estimation. Increase in RMSE relative to the oracle-lasso 
		as a function of the absolute distance ($|d|$) of the group-wise covariate means, under group one (top) and group two (bottom), and for cases A (left), B (center) and C (right). [Error bars indicate standard errors from 20 repetitions.]}
	\label{mse500}
\end{figure} 


\noindent\textbf{Estimation.}
Comparisons are made again with respect to the label-oracle-lasso; this time we consider the increase in root mean squared error (RMSE) resulting from the fact that the group labels are unknown. Here we also consider the \textit{pooled-lasso}; a ``naive'' model which does not take into account group structure. This allows us to investigate the effect of ignoring group structure under each case in Table \ref{sim2_param}; this is of particular interest in Case A where the common-location coefficients have opposite signs.

Results for the $p=500$ are summarized in Figure \ref{mse500}. We use standardized coefficients 
for the calculation of RMSE in order to have a common scale across simulations and cases. As expected, under Case A the pooled-lasso model performs poorly, while RJM methods provide overall better estimates than cluster-lasso. Under cases B and C, cluster-lasso and RJM which are over-parameterized (common-location effects are equal) perform more or less the same and are in general comparable to the pooled-lasso which is under-parameterized (due to the disjoint-location effects).
The $p=100$ and $p=250$ cases are shown in Appendix G (Figures \ref{mse100} and \ref{mse250}); results are in general similar with the difference that cluster-lasso performs better when $|d|\approx1$ and RJM-RLasso performs overall worse. Overall, our illustrations suggest that the RJM-FLasso is the most stable method, followed by RJM-NJ.

\medskip

\noindent\textbf{Selection of number of clusters.} In Appendix H we further consider the case where the number of clusters is not known \emph{a-priori} in simulation experiments which take into account all four cancer types. Results on cluster selection using the predictive approach described in Section \ref{prediction} are shown in Figure \ref{selection}.  

\subsection{Real cancer data example}
\label{sim4}

In this Section we consider a fully non-synthetic example, where both features and response are real empirical data. The general strategy is as follows. We use the TCGA data as introduced above, with gene expression levels treated as features. 
Specifically, we include all four cancer types used in the previous section, selecting via stratified random sampling  $n = 250$ samples in total. Stratified sampling ensures that the cancer-type proportions are preserved; specifically the dataset under consideration consists of 102 BRCA, 51 KIRC, 49 LUAD and 48 THCA observations (abbreviations as previously introduced). A total of $p=100$ gene expression levels (selected at random from all genes) are used in these experiments. As before, in the applications that follow the true labels (i.e., the cancer type indicator) are treated as latent and hence not used in analysis, but only to evaluate performance. As responses, we use one of the $p=100$ gene expression levels, with the remaining forming the feature set. This procedure has the advantage of allowing us to consider many different responses (genes) whilst entirely eschewing synthetic data generation. We first show a illustrative example using one particular gene as response and then show results from all responses considered.

\begin{figure}[h]
	\centering{}\includegraphics[scale=0.67]{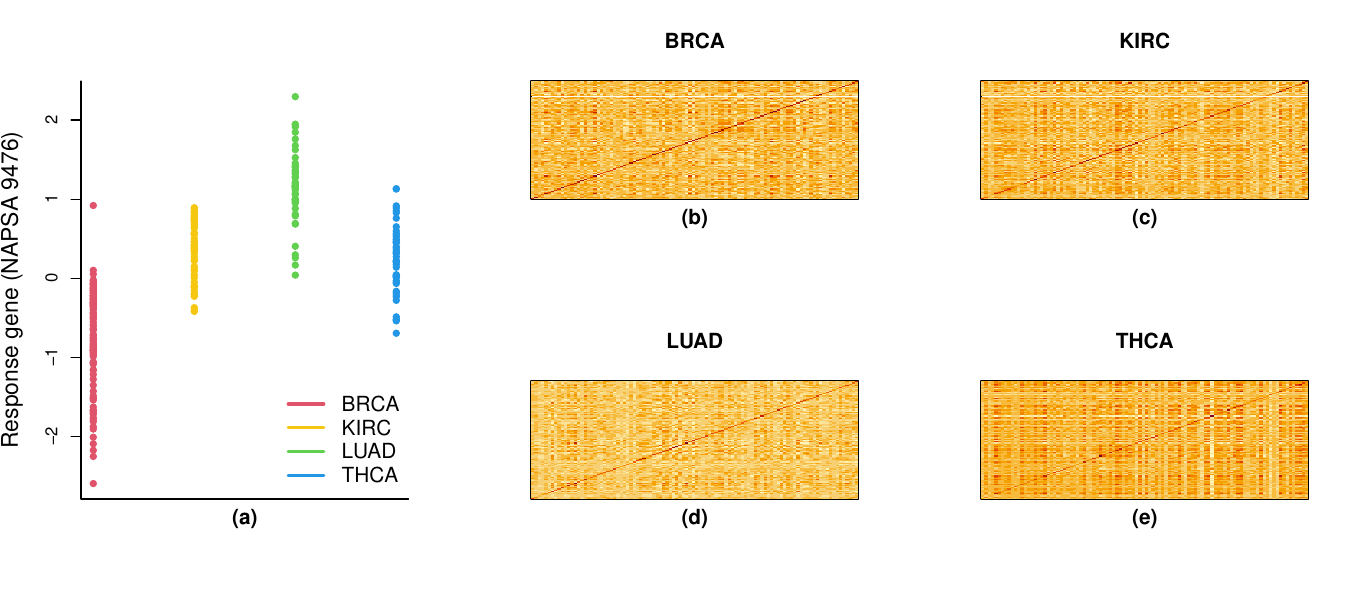}
	\caption{Real data example, data visualization for single, illustrative response. Plot (a) of response gene NAPSA  in the four cancer types, and heatmaps of feature covariances in (b) BRCA, (c) KIRC, (d) LUAD and (e) THCA cancer types.}
	\label{data}
\end{figure}

\medskip
\noindent 
{\bf Illustrative analysis for a particular response gene (NAPSA).} We consider the gene NAPSA (napsin A aspartic peptidase) (gene ID 9476) to illustrate the set-up. For this illustrative analysis we assume that the cancer types are given so that it is known \textit{a priori} that there are four classes. 
This particular gene is used to illustrate the approach as it is informative with respect to hidden group structure, as shown in Figure \ref{data} (left panel), but not to the extent of fully revealing the  class structure. Heatmaps of the sample covariance matrices of the remaining 99 genes under each cancer type are presented in Figure \ref{data} (panels in the right); these generally indicate slight differences in covariance structures.   

For evaluation of clustering performance we compare to the same methods as in Section \ref{sim1}, including again MoE as implemented in package \texttt{MoEclust}; however, for computational convenience, we use the option of a classification-EM algorithm, which is a faster but generally sub-optimal algorithm (although we note that initial tests suggested a gain in clustering performance for this dataset). We further consider $k$-medoids, fuzzy $c$-means (implemented via \texttt{par} and \texttt{fanny} in package \texttt{cluster}, respectively) and clustered support vector machines \citep[clustSVM, package \texttt{SwarmSVM};][]{Gu_Han_13}. Finally, for the purely cluster-oriented approaches ($k$-means/medoids, fuzzy $c$-means, hclust and mclust) we use as input the concatenated matrix containing the response and the predictor genes.  Table \ref{clustering} shows the resulting adjusted Rand index under each method (for $k$-means and clustSVM, which are highly sensitive to initialization, the values are averages from 100 runs). As seen, RJMs clearly outperform the other approaches.

\begin{table}[h]
	\centering
	\caption{Real data application, clustering performance. Adjusted Rand index values for the ten methods under consideration using gene NAPSA as response variable.}
	\begin{tabular}{lccccc}
		\hline
		\noalign{\vspace{0.1cm}}
		\multicolumn{6}{c}{\textbf{Methods and clustering performance}}\tabularnewline[\doublerulesep]
		\hline \\[-2ex]
		{Method} & $k$-means & $k$-medoids & fuzzy $c$-means  & hclust & clustSVM \tabularnewline
		{Adj. Rand Index}& 0.43 & 0.44 & 0.59 & 0.51 &  0.42 \tabularnewline
		\hline  \\[-2ex]
		{Method} &  mclust & MoEclust & RJM-NJ & RJM-FLasso & RJM-RLasso\tabularnewline
		{Adj. Rand Index}&  0.29 & 0.55 & 0.72 & 0.68 & 0.75\tabularnewline
		\hline
	\end{tabular}
	\label{clustering}
\end{table}

\begin{figure}[h]
	\centering{}\includegraphics[scale=0.75]{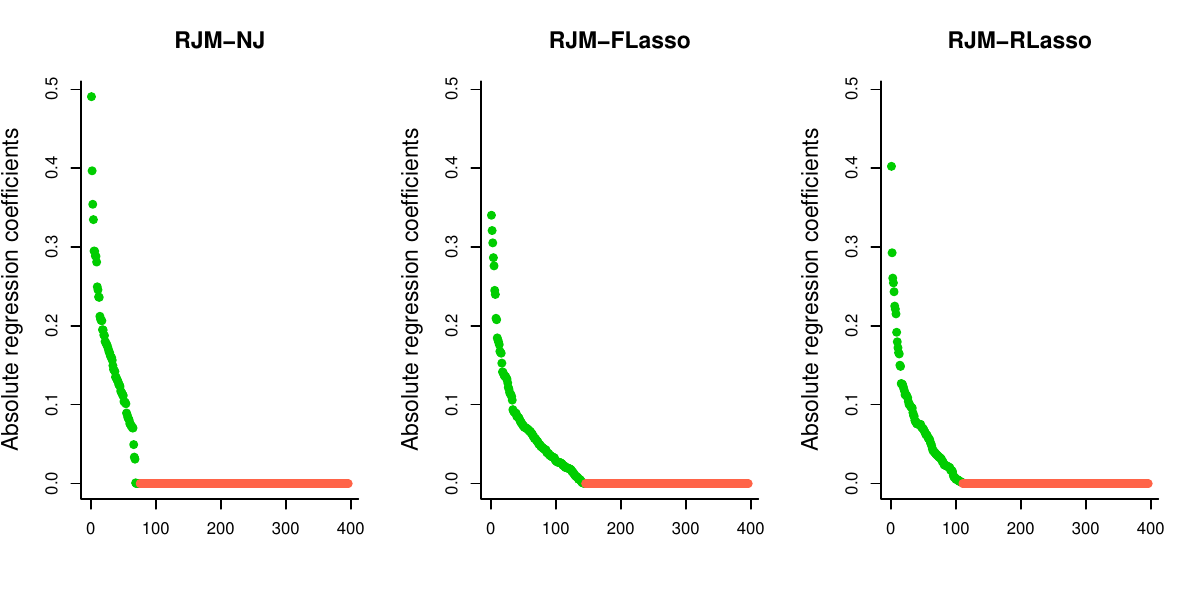}
	\caption{Real data application, regression performance. Absolute values of RJM regression coefficients ranked in decreasing order, using gene NAPSA as response variable. Green  points indicate non-zero coefficients; red points, coefficients that are set equal to zero.}
	\label{coef}
\end{figure}

Figure \ref{coef} shows the resulting regression coefficient estimates (396 in total given the four cancer types), from the three RJM variants, ranked in absolute value from highest to lowest (non-zero values in green; zero values in red). Consistent with the results presented in Section \ref{sim1}, we observe again that RJM-NJ results in the most parsimonious model (fewer than 100 predictors), followed by RJM-RLasso (around 100 predictors), while RJM-FLasso includes the most predictors (more than 100). 
We also observe that the lasso variants tend to shrink the coefficients of influential predictors more than RJM-NJ; this is also generally anticipated as the NJ prior has heavier tails in comparison to the Bayesian lasso prior.
Finally, the forward search of MoEclust included a few predictors in the gating networks, but resulted in entirely sparse expert networks as all  regression coefficients were set to zero.  

\noindent {\bf Performance over all responses.} 
Above we considered a specific gene to illustrate the key ideas; here we show results from all responses. That is, we consider in turn each of the genes as response, treating all others as features.  Thus, there are 100 problems considered  in total (each with the same four latent subgroups). In this case, the input for the purely clustering methods is the data matrix of the predictor variables. 

\begin{figure}[h]
	\centering{}\includegraphics[scale=0.38]{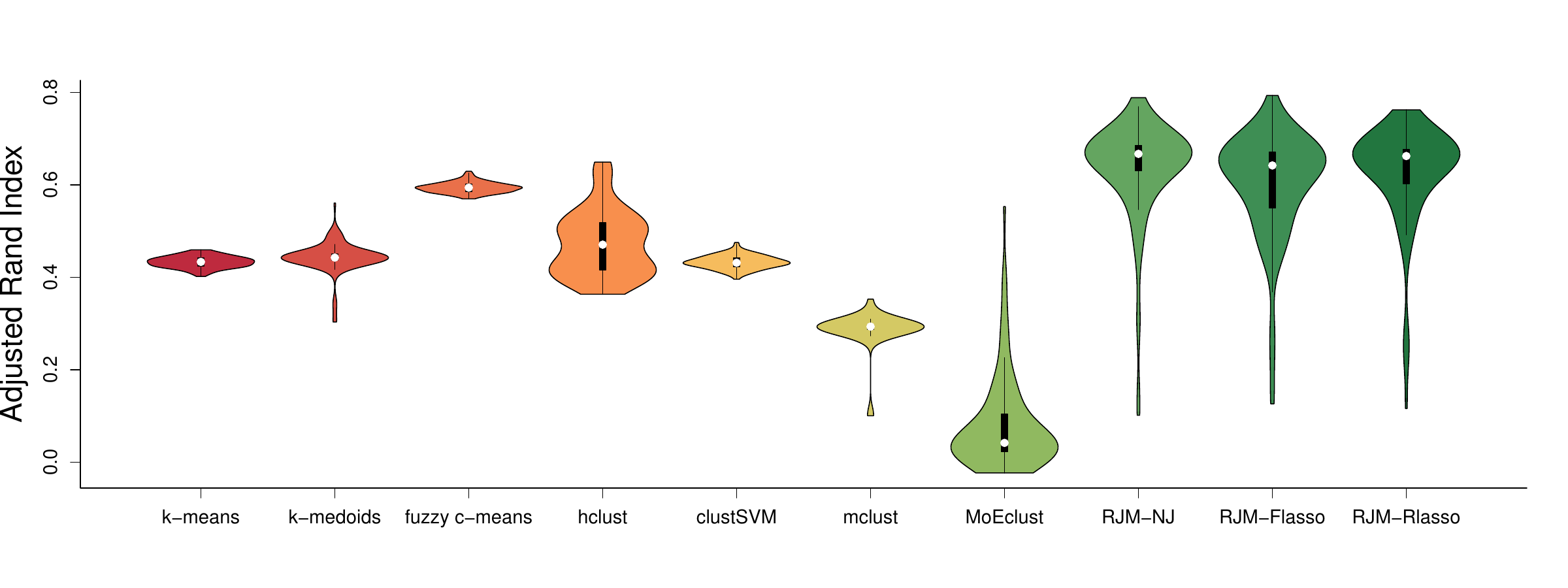}
	\caption{Real data application, clustering performance. Violin plots of adjusted Rand index values from TCGA-based experiments spanning a hundred different responses (see text for details).}
	\label{vioplot}
\end{figure}

Violin plots of the resulting adjusted Rand index values from the ten methods are presented in Figure \ref{vioplot}. 
These results, spanning one hundred different responses, support  the results seen above, as the three RJM variants consistently perform relatively well over most of the responses and the results are 
broadly in line with some of the results in Sections \ref{sim1} and \ref{sim2}.
In Appendix I, we further consider BIC-based model selection; as shown there RJMs select more frequently the correct number of groups in comparison to GMMs and MoEs.  

\subsection{Summary}
Broadly speaking
the RJM variants performed similarly to one another 
in terms of clustering/group assignment. 
In several situations they outperformed the other methods compared with, while, at the same time, they tended to remain competitive across the  range of scenarios tested.
Also, RJM can improve selection for the number of groups when this is not known at the outset.
There were some differences between the RJM variants with respect to regression modelling.
The NJ approach tended to perform well, in terms of variable selection and estimation, in sparse settings characterized by moderate to strong signals, while the lasso approaches yielded relatively denser models overall.

\section{Discussion}
\label{disc}

We introduced a class of regularized mixture models that jointly deal with sparse covariance structure and sparse regression in the context of latent groups. 
We showed that principled joint modeling of these two aspects leads to gains with respect to simpler decoupled or pooled strategies and that exploiting established $\ell_1$-penalized tools and related Bayesian approaches 
leads to practically applicable solutions. 
The RJM methods presented in this paper are implemented as an R package \texttt{regjmix}, available at \url{https://github.com/k-perrakis/regjmix}.
Future research directions include extensions 
to generalized linear models and mixed models.
Below we discuss some additional aspects and point to specific directions for future work.

\medskip
\noindent
{\bf Distribution shifts and shift-robust learning.} 
By accounting for data heterogeneity, RJMs help to  guard against (potentially severe) confounding of multivariate regression models 
by hidden group structure. 
This has interesting connections to distribution shifts in machine learning 
and shift-robust learning; see e.g. \cite{recht2018,heinze2019}. In particular, we think RJM would be a useful tool for shift-robust learning, since it could be used to block paired data $(X,Y)$ into distributionally non-identical groups which could in turn be used to train and test predictors in a shift-robust fashion, 
facilitating  shift-robust learning 
under unknown distributional regimes.

\medskip
\noindent
{\bf Choice of regularization.}
In this work we used the graphical lasso approach for covariance estimation, mainly motivated by certain biomedical applications where network models are of interest. However, the general RJM strategy could be used with other kinds of multivariate models (e.g. factor models). For the regression coefficients we considered: (i) the Bayesian lasso prior under two strategies (FLasso/RLasso)
and (ii) the NJ prior.
For practitioners seeking the closest analogue to the popular lasso approach based on cross-validation in the non-latent regression setting, we recommend FLasso. When it is preferable to use a shrinkage prior with heavier tails we recommend using NJ, which can be very effective in detecting sparsity patterns without over-shrinking large coefficients. 
In general, our main goal was to explore some of the available regularization options, 
but we note 
that the RJM model is fairly modular in the sense that other methods from the penalized likelihood  or the Bayesian literatures 
-- recent reviews provided by \cite{review_penalized} and \cite{review_shrinkage}, respectively -- can be used within the same framework. Of course, specifics will depend upon approach; for instance, the elastic-net \citep{zou_hastie_2005} and the adaptive lasso \citep{zou2006} are fairly easy to incorporate at present, while other methods such as the horseshoe estimator \citep{carvalho_etal_2010} require further investigation.

\medskip
\noindent
{\bf High-dimensional issues.} 
RJM remains effective when $p{>}n$, 
but a general 
issue when jointly modeling $(Y,X)$ is that for relatively large $p$ cluster allocation  will be mainly guided by $X$. In the empirical examples 
RJM outperformed \verb|mclust| (recall that without regularization RJM is equivalent to a GMM);
to provide some intuition about that let us consider the two regularization steps. The first  on the covariance matrix of $X$ 
can be viewed as $p$  lasso regressions \citep{Meinshausen2006}
that essentially discard non-influential relationships among features. The second
discards non-influential predictor effects on the response. Overall this sparsification may be viewed as a dimensionality-reduction, which mitigates 
over-emphasis on $X$. 
One idea for handling this issue as $p$ grows larger
is to consider explicit weighting of the effect of $X$,
e.g. by replacing the multivariate normal in \eqref{lik_x} with a density of the form
$\mathrm{N}_p(\xxi|\mk,\Sk)^{1/\delta}$, 
where the ``power-parameter'' ($\delta > 1$) 
would inflate the
covariance. 
While from a computational standpoint the proposed framework 
is scalable 
and can also handle the $p{>}n$ case, in very high dimensions
it would become computationally burdensome.
This can be potentially addressed via high-dimensional projections.
Finally, although our EM convergence result is general, there remain open theoretical questions concerning rates of convergence and optimality of the estimators themselves.

\section*{Acknowledgements}

We would like to thank Keefe Murphy for interesting discussions and for updating R package \texttt{MoEClust} in order to make it possible to implement the comparisons with mixtures-of-experts presented in this paper.  
We acknowledge support via the
German Bundesministerium f\"{u}r Bildung und Forschung (BMBF) project ``MechML", 
the Medical Research Council (programme number 
MC UU 00002/17)
and the National Institute for Health Research (NIHR) Cambridge Biomedical Research Centre.

%
%
%

\subsection*{Appendix A. Simulations for Section 1.2}
\label{AA}

The simulation results presented in Section \ref{motivation} are based on 20 repetitions where we consider two groups ($K=2$) with total $n=200$ and a balanced design, i.e. $n_k=100$ for $k = 1, 2$. The number of predictors is $p=10$. In each group only one  predictor ($\mathbf{x}^*_{k}$) has a non-zero coefficient ($\beta^*_k$), and this predictor is chosen randomly over the 20 repetitions. The three cases in Figures \ref{fig_intro1} and \ref{fig_intro} correspond to: (i) $\beta^*_1=\beta^*_2=0.5$ (plots on the left), (ii) $\beta^*_1=0.5, \beta^*_2=1$ (plots on the middle), and (iii) $\beta^*_1=0.5, \beta^*_2=1.5$ (plots on the right). 
The features are generated as
$\X_k\sim\N_{10}(\mm_k, 0.5\Ip)$. That is, the two feature groups share the same diagonal covariance structure, but we let the mean vectors to vary. Specifically, the first mean vector is always $\mm_1=(0,\dots,0)^T$, while for the second mean vector we consider again three cases (which lead to the variability with respect to the $x$-axis of the plots in Figure \ref{fig_intro}); namely, (i) $\mm_2=(0,\dots,0)^T$, (ii) $\mm_2=(0.5,\dots,0.5)^T$ and (iii) $\mm_2=(1,\dots,1)^T$.
The response of each group is generated as
$
\by_k\sim \N_{n_k}\big(\mathbf{x}^*_{k}\beta^*_{k},~ \sigma_k^2\mathbf{I}_{n_k}\big) 
$, where $\sigma_k^2 = \mathrm{Var}(\mathbf{x}^*_{k}\beta^*_{k})/5$ for $k = 1,2$. 
The regularized joint mixture model is based on the normal-Jeffreys prior discussed in Section \ref{regression}.
For the implementation of the Gaussian mixture and mixture-of-experts models we used \verb|R| packages \texttt{mclust} \citep{mclust} and \texttt{MoEClust} \citep{murphy_murphy2020}, respectively, using the default model-search options, selecting the BIC-optimal model.

\subsection*{Appendix B. Justification for the RLasso Pareto prior}
\label{AAA}



We generally want the Pareto prior to be such that it will not penalize the regression coefficients asymptotically. 
Under the prior in \eqref{prior_lambda} the mode of $\lambda_k$
is $a_n$, while the prior mean is given by
\begin{equation*}
\mathbb{E}(\lambda_k)=a_n\displaystyle\frac{b_n}{b_n-1},
\end{equation*}
for $b_n > 1$ and the prior variance by
\begin{equation*}
\mathrm{Var}(\lambda_k)= a_n^2\displaystyle\frac{b_n}{(b_n-1)^2(b_n-2)}
\end{equation*}
for $b_n > 2$.
Given that $a_n\rightarrow 0$ as $n\rightarrow \infty$, in order to meet our requirement we want $b_n \rightarrow C>2$ as $n\rightarrow\infty$; to that end, we specify 
$
b_n=(p-1)- c\sqrt{2K\log p/n},
$
for $c\in(0,1]$.
Explicit specification of $a_n$ is not required as it does not affect the posterior mode; any decreasing function of $n$ (subject to $a_n>0$) will satisfy the recuirement. As for $c$, 
we recommend setting it equal to $\min\{\sqrt{2p/3n},1\}$ as a default option.

\subsection*{Appendix C. Objective function under the NJ prior}
\label{AppA}

The hierarchical form of the NJ prior is $\bbk|\Laj \sim \N_p(\boldsymbol{0},\Laj)$, where $\Laj=\diag(s_{k1},\dots,s_{kp})$ assuming \textit{latent} $\lj$ with $\pi(\lj)\propto \lj^{-1}$ for $k=1,\dots,K$ and $j=1,\dots,p$. 
The conditional distribution of any $s$ (dropping momentarily subscripts $k,j$ for simplicity) is 
\begin{equation*}
p(s|\beta)=\frac{q(s)}{\displaystyle\int q(s) \mathrm{d}s}, ~ \mbox{with} ~ \displaystyle q(s) = p(\beta|s)s^{-1} ~ \mbox{and} ~ \displaystyle\int q(s) \mathrm{d}s = |\beta|^{-1}.
\end{equation*}
Given this, it follows that 
\begin{equation}
\mathbb{E}_{s|\beta}[s^{-1}] = \int s^{-1}p(s|\beta)\mathrm{d}s={\beta^{-2}}
\label{A1}
\end{equation} 
This result will be needed in the derivation of the E-step below.
The joint prior of $\bbk$ and $\sk$ is
$p(\bb,\s|\La)
= p(\bb|\La)p(\s)
\propto
\prod_{k=1}^{K}
\exp\big(
-\frac{1}{2}\bbk^{T}\invLaj\bbk
\big)\frac{1}{\sk}.$
The objective under the NJ prior presented in Eq. \eqref{NJobj} in the main paper, is derived as follows.
\begin{align}
Q^{\scaleto{Y}{4pt}}_{\scaleto{\mathrm{NJ}}{4pt}}(\boldsymbol{\theta}^{\scaleto{Y}{4pt}}|\boldsymbol{\theta}^{{\scaleto{Y}{4pt}}(t)}) = & 
\mathbb{E}_{\mathbf{z},\La|\by,\X,\boldsymbol{\theta}^{{\scaleto{Y}{4pt}}(t)}}\big[\log f(\by|\X,\mathbf{z},\al,\bb,\s)
+
\log\pi(\bb|\La) + \log\pi(\s)
\big] \nonumber
\\
=&
\mathbb{E}_{\mathbf{z}|\by,\X,\boldsymbol{\theta}^{{\scaleto{Y}{4pt}}(t)}}\big[\log f(\by|\X,\mathbf{z},\al,\bb,\s)\big]
+
\mathbb{E}_{\La|\bb^{(t)}}\big[\log\pi(\bb|\La)
\big]+\log\pi(\s) \nonumber
\\
=&
\sum_i\mathbb{E}_{\mathbf{z}|\by,\X,\boldsymbol{\theta}^{{\scaleto{Y}{4pt}}(t)}}\big[\log f(y_i|\xxi,z_i,\al,\bb,\s)\big]
+
\mathbb{E}_{\La|\bb^{(t)}}\big[\log\pi(\bb|\La)
\big]+\log\pi(\s) \nonumber
\\
=& \sum_{i=1}^{n} \sum_{k=1}^{K}  m^{(t)}_{ki}\Bigg\{
-\frac{1}{2\sk}(y_i-\ak-\xxti\bbk)^2
-\frac{1}{2}\log\sk\Bigg\} \label{A2}
\\
&
+
\sum_{k=1}^{K} \Bigg\{ 
\mathbb{E}_{\La_k|\bbk^{(t)}}
\Bigg[
-\frac{1}{2}\bbk^{T}\invLaj\bbk
\Bigg]
\Bigg\} - 
\sum_{k=1}^{K} \Big\{
\log\sk
\Big\} \nonumber
\end{align}
\begin{align}
=& 
\sum_{k=1}^{K}  \Bigg\{ 
-\frac{1}{2\sk} (\by-\ak\mathbf{1}_{n}-\X\bbk)^{T}\M_k^{(t)}(\by-\ak\mathbf{1}_{n}-\X\bbk)
-\frac{n_k^{(t)}}{2}\log\sk
\Bigg\} \nonumber
\\
&
+
\sum_{k=1}^{K} \Bigg\{ 
-\frac{1}{2}\bbk^{T}
\mathbb{E}_{\La_k|\bbk^{(t)}}\big[
\invLaj
\big]\bbk
\Bigg\} - 
\sum_{k=1}^{K} \Big\{
\log\sk
\Big\} \label{A3}
\\
=& 
\sum_{k=1}^{K}  \Bigg\{ 
-\frac{1}{2\sk} (\by-\ak\mathbf{1}_{n}-\X\bbk)^{T}\M_k^{(t)}(\by-\ak\mathbf{1}_{n}-\X\bbk)
-\frac{n_k^{(t)}+2}{2}\log\sk
\Bigg\} \nonumber
\\
&
+
\sum_{k=1}^{K} \Bigg\{ 
-\frac{1}{2}\bbk^{T}
\V_k^{(t)}\bbk
\Bigg\} \label{A4}
\\
=& 
-\frac{1}{2}\sum_{k=1}^{K}  \Bigg\{ 
\frac{(\by-\ak\mathbf{1}_{n}-\X\bbk)^{T}\M_k^{(t)}(\by-\ak\mathbf{1}_{n}-\X\bbk)}{\sk}+\bbk^{T}\V^{(t)}_k\bbk \nonumber
\\
&
~~~~~~~~~~~~~+(n_k^{(t)}+2)\log\sk
\Bigg\}, \nonumber
\end{align}
where $m_{ki}^{(t)}$ appearing in \eqref{A2} is given in \eqref{member} in the main paper, $\M_k^{(t)}=\diag(\mathbf{m}_k^{(t)})$ with $\mathbf{m}_k^{(t)}=(m_{k1}^{(t)},\dots,m_{kn}^{(t)})^T$ and $\V_k^{(t)} = \diag\Big(1/\beta_{k1}^{2(t)},\dots,1/\beta_{kp}^{2(t)}\Big)$. The transition from \eqref{A3} to \eqref{A4} is due to \eqref{A1}.

\subsection*{Appendix D. Details and implementation of the EM}
\label{AppImplementation}

For the graphical lasso optimization in \eqref{omega} in the main paper we use the efficient R package \verb|glassoFast| \citep{glassofast}. For the lasso optimizations in \eqref{phi} we use \verb|glmnet| \citep{glmnet} with penalty equal to $\lambda_{k}^{(t+1)}/n_k^{(t)}$. 

We initialize the algorithm via a simple clustering of the data. For this we use R package \verb|mclust|. Through the resulting group assignments we obtain initial estimates $\boldsymbol{\theta}^{\scaleto{X(0)}{6pt}}_k$ and $\boldsymbol{\theta}_k^{\scaleto{Y(0)}{6pt}}$. In order to initiate EMs from different starting points we add random perturbations  to $\mk^{\scaleto{(0)}{6pt}}$, $\bbk^{\scaleto{(0)}{6pt}}$ and $\sigma_k^{\scaleto{2(0)}{6pt}}$ and positive random perturbations to the diagonal elements of $\Sk^{\scaleto{(0)}{6pt}}$.
The multiple EMs can be easily run in parallel. As a default option we use ten EM starts.

For the termination of the algorithm we use a combination of two criteria that are commonly used in practice. The first is to simply set a maximum number $(T)$ of EM iterations. Empirical results suggest that the option $T=20$ is sufficient. The second criterion takes into account the relative change in the objective function in \eqref{obj}; namely, the algorithm is stopped when
\begin{equation*}
\Bigg\vert
\frac
{Q(\thetab,\boldsymbol{\tau},\boldsymbol{\lambda}|\thetab^{(t)},\boldsymbol{\tau}^{(t)},\boldsymbol{\lambda}^{(t)})}
{Q(\thetab,\boldsymbol{\tau},\boldsymbol{\lambda}|\thetab^{(t-1)},\boldsymbol{\tau}^{(t-1)},\boldsymbol{\lambda}^{(t-1)})}
-1
\Bigg
\vert
\leq \epsilon
\end{equation*}
using as default option $\epsilon=10^{-6}$. Moreover, the algorithm is stopped, and results are discarded, when the sample size of a certain group becomes prohibitively small for estimation. We define this criterion as a function of total sample size and the number of groups. Specifically, we terminate if $\min_k n_k^{(t)}\leq n/(10K)$.

\subsection*{Appendix E. Proof of Theorem 2} \label{AppTheory}
We need to prove that with our model and under the assumptions of Theorem \ref{thm:convergence_ECM_RJM}, all the hypotheses of Theorem 3 of \cite{Meng_Rubind1993} (Theorem \ref{thm:convergence_ECM}) are met.

As discussed throughout section \ref{sect:main_convergence_result} in the main paper, under the RJM model, the following conditions are sufficient to verify all the hypotheses of Theorem 4.1 except for hypothesis 6: the penalty is continuous, differentiable, such that each of the block maximisation is unique and $L(\boldsymbol\xi)$ infinite on the border of $\Xi$. The first three conditions are already assumptions of our Theorem. Hence, it remains only to be shown that $L(\boldsymbol\xi)$ infinite on the border of $\Xi$ and that hypothesis (6) is met. Both are very similar properties, we prove both them together. For this task, we make use of the ``penalty lower bound assumption" of our Theorem, recalled in Eq.~\eqref{eq:sufficient_penalty_lower_bound}.
\begin{equation}\label{eq:sufficient_penalty_lower_bound}
\mathrm{pen}(\boldsymbol{\xi})\geq \delta \sum_{k=1}^{K} \parent{\log \tau_k^{-1} + \norm{\mu_k} + \norm{\Om_k} + \log \det{\Om_k^{-1}}  +  f_{\lambda}(\lambda_k) + \rho_k + \log\rho_k^{-1} +   \det{\chi_k} + \norm{\f_k}}\, .
\end{equation}
To begin with, we have:
\begin{equation*}
\begin{split}
L(\boldsymbol\xi) &= \sum_{i=1}^{n} \log \sum_{k=1}^{K} \exp\Bigg( -\frac{1}{2} \bigg( (y_i \rho_k - \chi_k - \xii^T \f_k)^2 -2 \log \rho_k\\
&\hspace{4.2cm}+ \norm{\xii-\m_k}_{\Om_k}^2 - \log \det{\Om_k} \\
&\hspace{4.2cm} -2\log\tau_k + (p+1) \log 2\pi + \frac{2}{n} \mathrm{pen}(\boldsymbol\xi) \bigg)\Bigg)\, .\\
\end{split}
\end{equation*}
Let 
\begin{equation*}
f_{i, k}(\boldsymbol\xi) = (y_i \rho_k - \chi_k - \xii^T \f_k)^2 -2 \log \rho_k + \norm{\xii-\m_k}_{\Om_k}^2 - \log \det{\Om_k}  -2\log\tau_k + \frac{2}{n} \mathrm{pen}(\boldsymbol\xi) \, .
\end{equation*}
Such that
\begin{equation}\label{eq:proof_likelihood_reformulation}
L(\boldsymbol\xi) = -\frac{n(p+1)\log2\pi}{2}+\sum_{i=1}^{n} \log \sum_{k=1}^{K} \exp\parent{ -\frac{1}{2} f_{i, k}(\boldsymbol\xi)} \, .
\end{equation}
From Eq.~\eqref{eq:sufficient_penalty_lower_bound} and the fact that $ (y_i \rho_k - \chi_k - \xii^T \f_k)^2 \geq 0$ and $\norm{\xii-\m_k}_{\Om_k}^2 \geq 0$, we have:
\begin{equation}\label{eq:proof_lower_bound}
\begin{split}
f_{i, k}(\boldsymbol\xi) &\geq \frac{2}{n}\delta \sum_{l=1}^{K} \Big(-(1+\frac{n}{\delta}\mathds{1}_{l=k})\log \tau_l + \norm{\mu_l} + \norm{\Om_l} - (1+\frac{n}{2\delta}\mathds{1}_{l=k})\log \det{\Om_l}\\
&\hspace{2cm}+ f_{\lambda}(\lambda_l) + \rho_l - (1+\frac{n}{\delta}\mathds{1}_{l=k})\log\rho_l +   \det{\chi_l} + \norm{\f_l}\Big) \\
&= \frac{2}{n}\delta \sum_{l=1}^{K} \Big(f_{k, l}^{\boldsymbol\tau}(\tau_l)+f_{k,l}^{\m}(\m_l)+f_{k, l}^{\Om}(\Om_l) + f_{k, l}^{\boldsymbol\lambda}(\lambda_l) + f_{k, l}^{\boldsymbol\rho}(\rho_l) + f_{k, l}^{\boldsymbol\chi}(\chi_l) + f_{k, l}^{\f}(\f_l)\Big)  \, .
\end{split}
\end{equation}
Where:
\begin{equation}\label{eq:proof_parameter_wise_penalties}
\begin{split}
f_{k, l}^{\boldsymbol\tau}(\tau_l) &:=-(1+\frac{n}{\delta}\mathds{1}_{l=k})\log \tau_l\, ,\\
f_{k,l}^{\m}(\m_l) &:=\norm{\m_l}\, ,\\
f_{k, l}^{\Om}(\Om_l) &:=\norm{\Om_l} - (1+\frac{n}{2\delta}\mathds{1}_{l=k})\log \det{\Om_l}\, ,\\
f_{k, l}^{\boldsymbol\lambda}(\lambda_l) &:= f_{\lambda}(\lambda_l)\, ,\\
f_{k, l}^{\boldsymbol\rho}(\rho_l) &:=\rho_l - (1+\frac{n}{\delta}\mathds{1}_{l=k})\log\rho_l\, ,\\
f_{k, l}^{\boldsymbol\chi}(\chi_l) &:=\det{\chi_l}\, , \\
f_{k, l}^{\f}(\f_l) &:=\norm{\f_l} \, .
\end{split}
\end{equation}
The dependency on $k, l$ is denoted in the indices of all these functions for the sake of uniformity, although only $f_{k, l}^{\boldsymbol\tau}, f_{k, l}^{\Om}$ and $f_{k, l}^{\boldsymbol\rho}$ actually depend on $k$ and $l$. We recall that, with $a>0$, the function $x \mapsto x - a \log x$, is lower bounded on $\R_+^*$, and converges towards $+\infty$ both when $x\to0$ and when $x\to+\infty$. To analyse $f_{k, l}^{\Om}$, it is convenient to consider the nuclear norm for $\norm{\Om_k}$ and rewrite the whole as: $f_{k, l}^{\Om}(\Om_l) = \sum_j \psi_{l, j} - (1+\frac{n}{2\delta}\mathds{1}_{l=k}) \log \psi_{l, j}$, with $\brace{\psi_{l, j}}_{j=1}^p$ the eigenvalues of $\Om_l$. 

With these observations at hand, note that all the functions in \eqref{eq:proof_parameter_wise_penalties} can be lower bounded by the same constant $c>-\infty$, valid for all values of $k$ and $l$. They also all converge towards $+\infty$ on the boundary of their respective sets of definition. 

For $m>0$, we define $\Xi_m$ as the compact subset of $\Xi$ such that $\xi \in \Xi_m \iff \xi \in \Xi$ and $\forall k = 1, ..., K:$
\begin{equation} \label{eq:proof_compact_definition}
\begin{split}
\tau_k \geq \frac{1}{m} \, ,\\
\norm{\m_k} \leq m \,, \\
\psi_{\text{max}}(\Om_k)  \leq m \,, \\
\psi_{\text{min}}(\Om_k) \geq \frac{1}{m} \,, \\
\frac{1}{m} \leq \lambda_k \leq m \, ,\\
\frac{1}{m} \leq \rho_k \leq m \, ,\\
\det{\chi_k} \leq m \,, \\
\norm{\f_k} \leq m \, .\\
\end{split}
\end{equation}
It is clear that $\forall m>0, \; \Xi_m \subseteq \Xi^{\circ}$.

With all these objects defined, we can finish the proof. For any real number $A>-\infty$, let us show that there exists $M>0$ such that $\forall\xi \in \Xi\setminus\Xi_M, \; L(\xi)<A$. First consider the following: for any real number $B>-\infty$, there exists a $m_B>0$ such that, for all $k$ and $l$:
\vspace{1em}
\begin{equation}
\begin{split}
&\text{if } \tau_l < \frac{1}{m_B} \text{, then } f_{k, l}^{\boldsymbol\tau}(\tau_l) > B\, ,\\
&\text{if } \norm{\m_l} > m_B \text{, then } f_{k,l}^{\m}(\m_l) > B\,, \\
&\text{if }\psi_{\text{max}}(\Om_l)  > m_B \text{, then } f_{k, l}^{\Om}(\Om_l) > B\,, \\
&\text{if }\psi_{\text{min}}(\Om_l) < \frac{1}{m_B} \text{, then } f_{k, l}^{\Om}(\Om_l) > B\,, \\
&\text{if } \lambda_l<\frac{1}{m_B} \text{ or }\lambda_l> m_B \text{, then } f_{k, l}^{\boldsymbol\lambda}(\lambda_l) > B\, ,\\
&\text{if } \rho_l<\frac{1}{m_B} \text{ or }\rho_l> m_B \text{, then } f_{k, l}^{\boldsymbol\rho}(\rho_l) > B\, ,\\
&\text{if }\det{\chi_l} > m_B \text{, then } f_{k, l}^{\boldsymbol\chi}(\chi_l) > B\,, \\
&\text{if }\norm{\f_l} > m_B \text{, then } f_{k, l}^{\f}(\f_l) > B\, .\\
\end{split}
\end{equation}
If $\xi \in \Xi\setminus\Xi_{m_B}$, then by definition of the sets $\Xi_m$ \eqref{eq:proof_compact_definition}, there exist at least one $l \in \{1, ..., K\}$ such that at least one of the above scenarios is realised. By injecting the resulting lower bound into the inequality \eqref{eq:proof_lower_bound}, we get:
\begin{equation*}
\forall k, \quad f_{i, k}(\boldsymbol\xi) > \frac{2}{n}\delta (B + (7K-1)c)  \, .
\end{equation*}
Then, from Eq.~\eqref{eq:proof_likelihood_reformulation}:
\begin{equation*}
\begin{split}
L(\boldsymbol\xi) &= -\frac{n(p+1)\log2\pi}{2}+\sum_{i=1}^{n} \log \sum_{k=1}^{K} \exp\parent{ -\frac{1}{2} f_{i, k}(\boldsymbol\xi)} \\
&< -\frac{n(p+1)\log2\pi}{2}+\sum_{i=1}^{n} \log \sum_{k=1}^{K} \exp\parent{ -\frac{1}{n}\delta (B + (7K-1)c) }\\
&= -\frac{n(p+1)\log2\pi}{2}+ n \log K - \delta (B + (7K-1)c) \, .
\end{split}
\end{equation*}
Since $\delta>0$, then there exists $B_A>0$ such that for all $B\geq B_A$:
\begin{equation*}
-\frac{n(p+1)\log2\pi}{2}+ n \log K - \delta (B + (7K-1)c) <A \, .
\end{equation*}
As a consequence, $M:=m_{B_A}$ is such that $\forall\boldsymbol\xi \in \Xi\setminus\Xi_M, \; L(\boldsymbol\xi)<A$. In other words $\brace{\boldsymbol\xi \in \Xi | L(\boldsymbol\xi) \geq A} \subseteq \Xi_M$. 

We have proven that for any $A>-\infty$, there exists $M>0$ such that $\brace{\boldsymbol\xi \in \Xi | L(\boldsymbol\xi) \geq A} \subseteq \Xi_M$. Since the $\Xi_m$ are compacts, this means that the closed set $\brace{\boldsymbol\xi \in \Xi | L(\boldsymbol\xi) \geq A}$ is also a compact, hence hypothesis (6) of \cite{wu1983convergence} is verified. Moreover, since $\Xi_m \subseteq \Xi^{\circ}$, this means that the log-likelihood goes to $-\infty$ on the border of $\Xi$. Hence no EM sequence will take values on the border, hence we can safely consider that $\Xi=\Xi^{\circ}$. With these last two hypotheses verified, we can apply Theorem 3 of \cite{Meng_Rubind1993} and benefit from the convergence guarantees.

\subsection*{Appendix F. Further results from Section 5.1} 
\label{AppB}

\begin{table}[h!]
	\centering
	\caption{First simulation. Intercept and slope parameter values for the two groups under the three cases illustrated in Figure 1 in the main paper.}
	\begin{tabular}{cccc}
		\hline
		\noalign{\vspace{0.1cm}}
		\textbf{Case} & \textbf{Group} & \textbf{Intercept} & \textbf{Slope}\tabularnewline[\doublerulesep]
		\hline 
		\multirow{2}{*}{A} & 1 & 0 & \color{white}-\color{black}1\tabularnewline
		& 2 & 0 & -1\tabularnewline
		\hline 
		\multirow{2}{*}{B} & 1 & 0 & \color{white}-\color{black}1\tabularnewline
		& 2 & 1 & \color{white}-\color{black}1\tabularnewline
		\hline 
		\multirow{2}{*}{C} & 1 & 0 & \color{white}-\color{black}1\tabularnewline
		& 2 & 0 & \color{white}-\color{black}1\tabularnewline
		\hline
	\end{tabular}
	\label{sim1_param}
\end{table}

\begin{figure}[h]
	\centering{}\includegraphics[height=10cm,width=\textwidth]{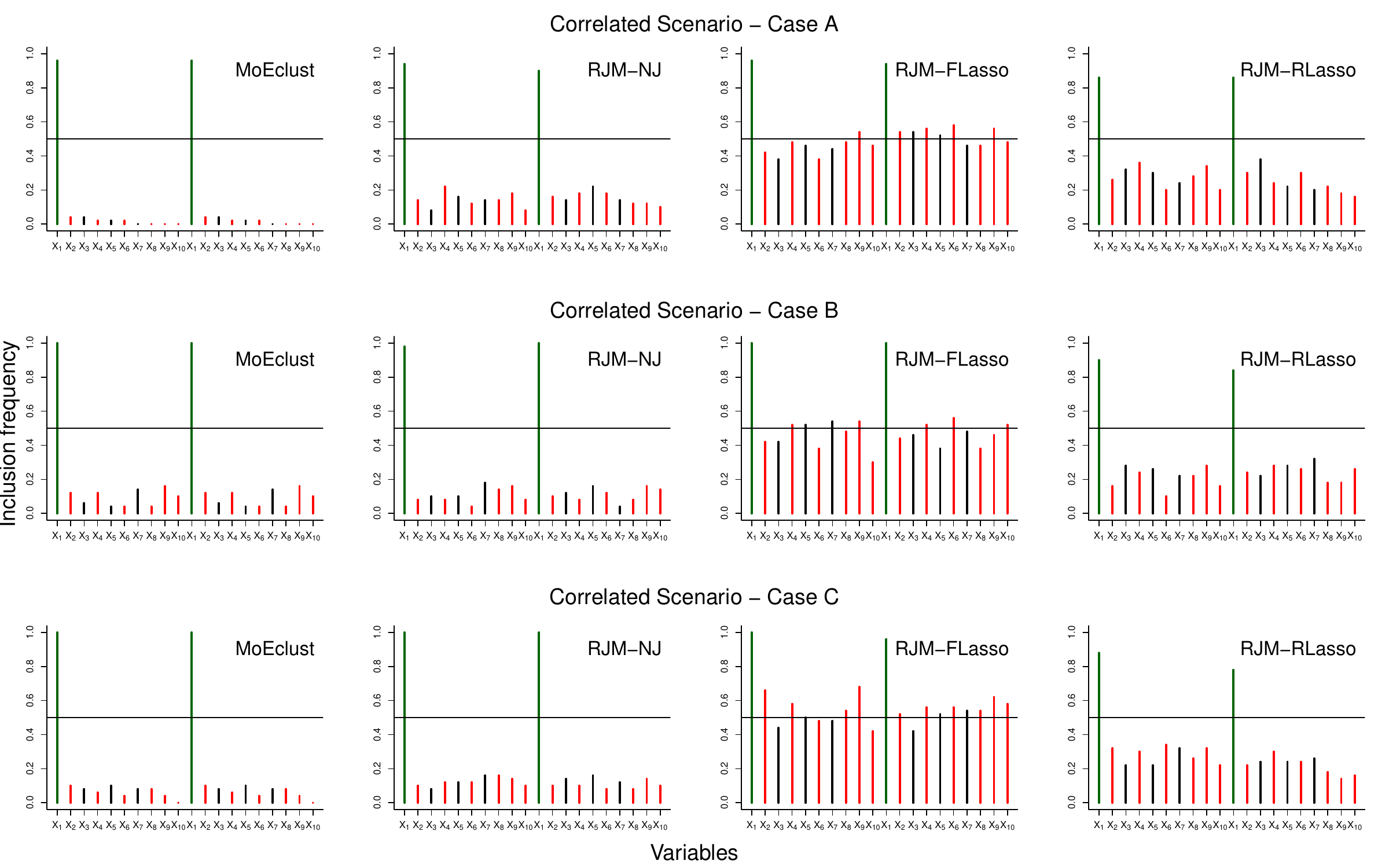}
	\caption{First simulation, correlated scenario. Variable inclusion frequencies (under 50 repetitions) for signal variables (in green), correlated noise variables (in black) and uncorrelated noise variables (in red) for regression cases A, B and C. Horizontal black lines correspond to a frequency of 0.5.}
	\label{inc.corr}
\end{figure}
\clearpage
\begin{figure}
	\centering{}\includegraphics[height=6cm,width=\textwidth]{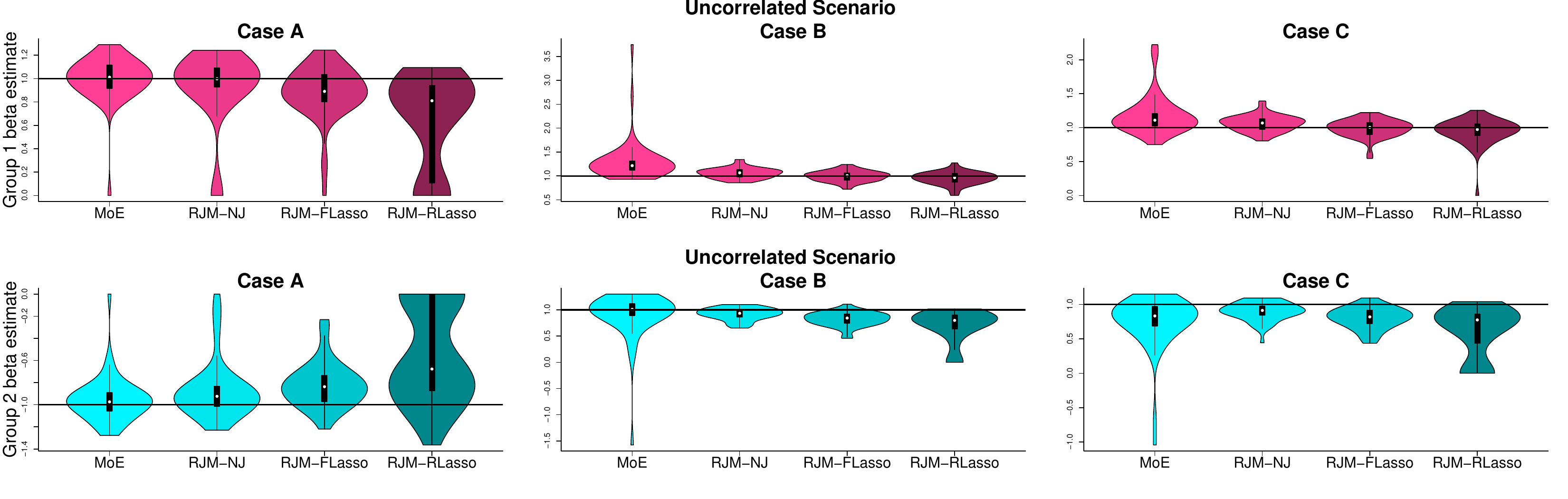}
	\caption{First simulation, uncorrelated scenario. Violin plots of slope MoEclust and RJM estimates (from 50 repetitions) for cases A, B and C. Horizontal black lines correspond to the true slopes.}
	\label{slope.diag}
\end{figure}
\begin{figure}
	\centering{}\includegraphics[height=6cm,width=\textwidth]{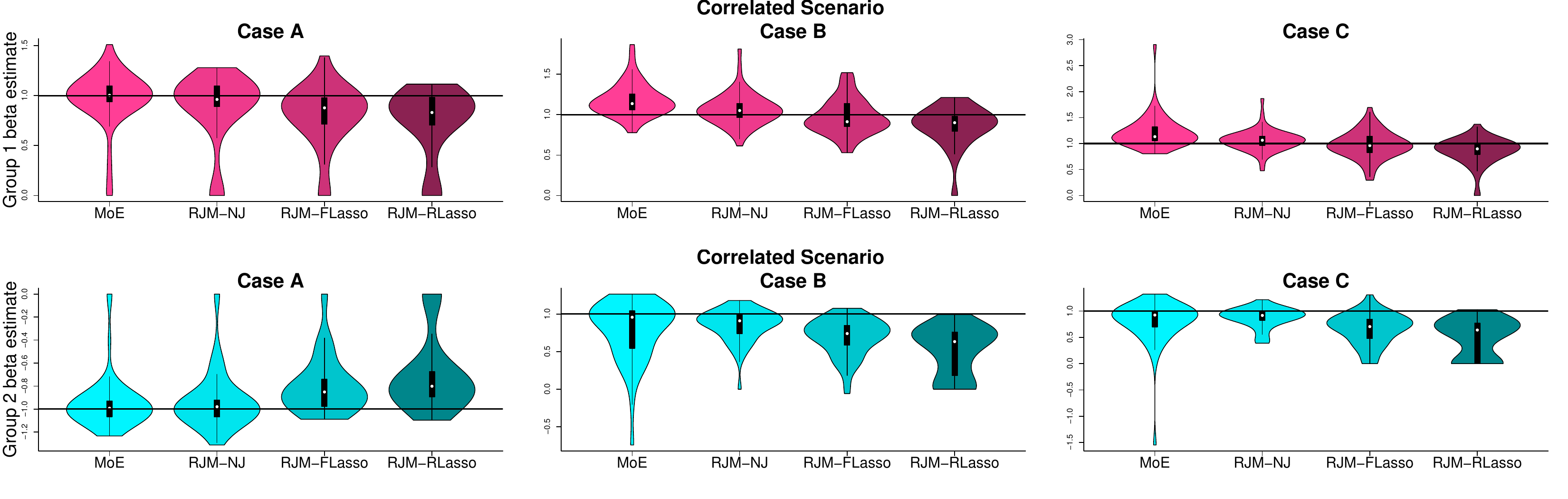}
	\caption{First simulation, correlated scenario. Violin plots of slope MoEclust and RJM estimates (from 50 repetitions) for cases A, B and C. Horizontal black lines correspond to the true slopes.}
	\label{slope.corr}
\end{figure}
\clearpage
\begin{figure}[h!]
	\centering{}\includegraphics[height=6cm,width=\textwidth]{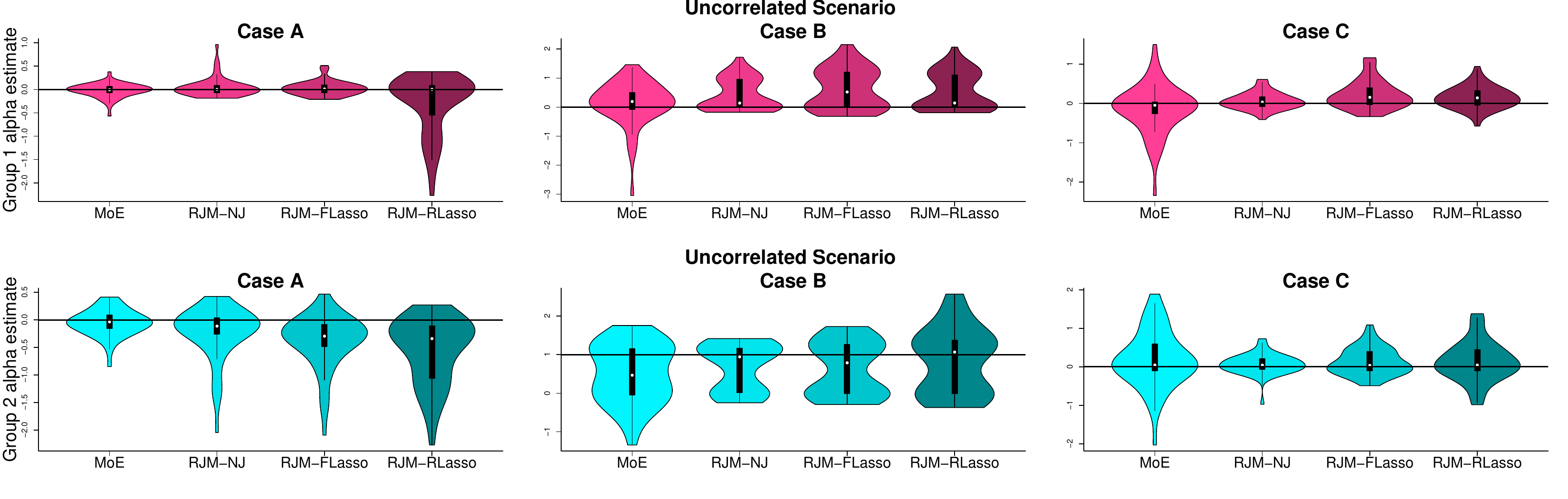}
	\caption{First simulation, uncorrelated scenario. Violin plots of intercept MoEclust and RJM estimates (from 50 repetitions) for cases A, B and C. Horizontal black lines correspond to the real intercepts.}
	\label{intercept.diag}
\end{figure}
\begin{figure}[h!]
	\centering{}\includegraphics[height=6cm,width=\textwidth]{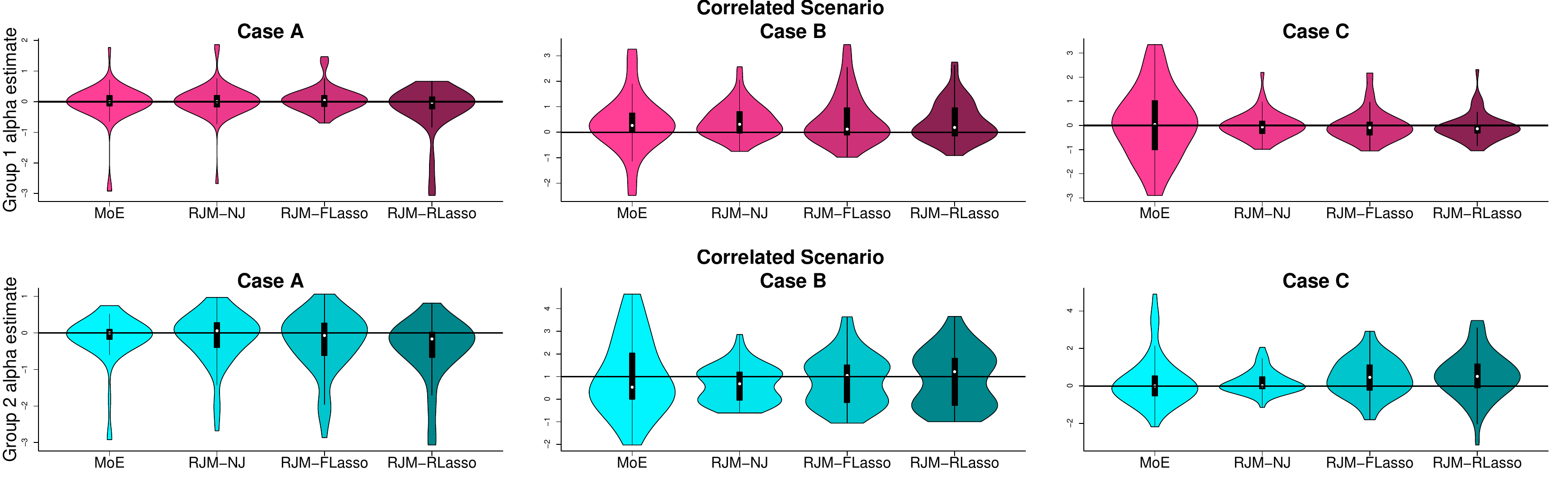}
	\caption{First simulation, correlated scenario. Violin plots of intercept MoEclust and RJM estimates (from 50 repetitions) for cases A, B and C. Horizontal black lines correspond to the real intercepts.}
	\label{intercept.corr}
\end{figure}

\subsection*{Appendix G. Further results from Section 5.2} 
\label{AppC}

\begin{figure}[H]
	\centering{}\includegraphics[height=7cm,width=\textwidth]{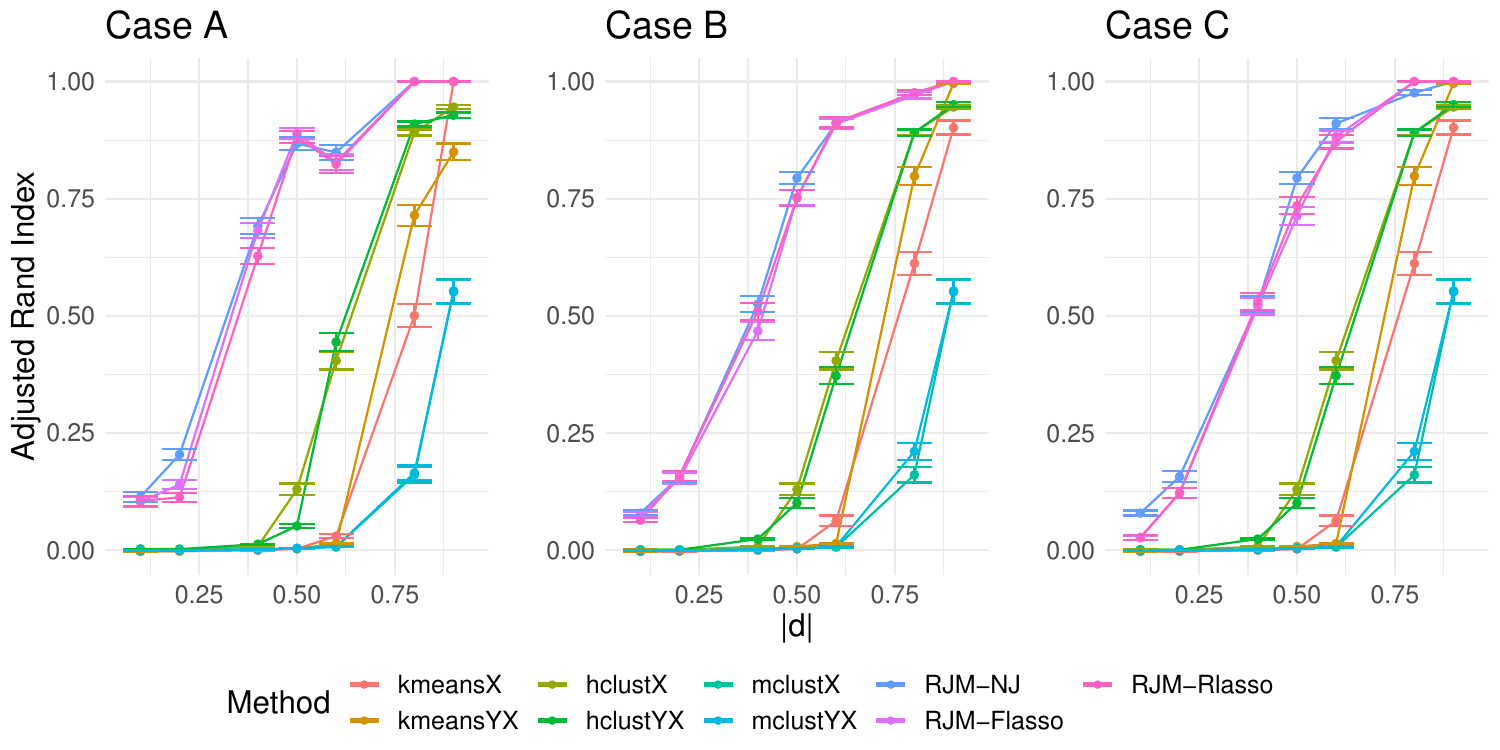}
	\caption{Second simulation, $p=100$, group assignment. Average adjusted Rand Index 
		as a function of the absolute distance ($|d|$) of the group-wise covariate means, for cases A (left), B (center) and C (right). [Error bars indicate standard errors from 20 repetitions.]}
	\label{rand100}
\end{figure}

\begin{figure}[H]
	\centering{}\includegraphics[height=7cm,width=\textwidth]{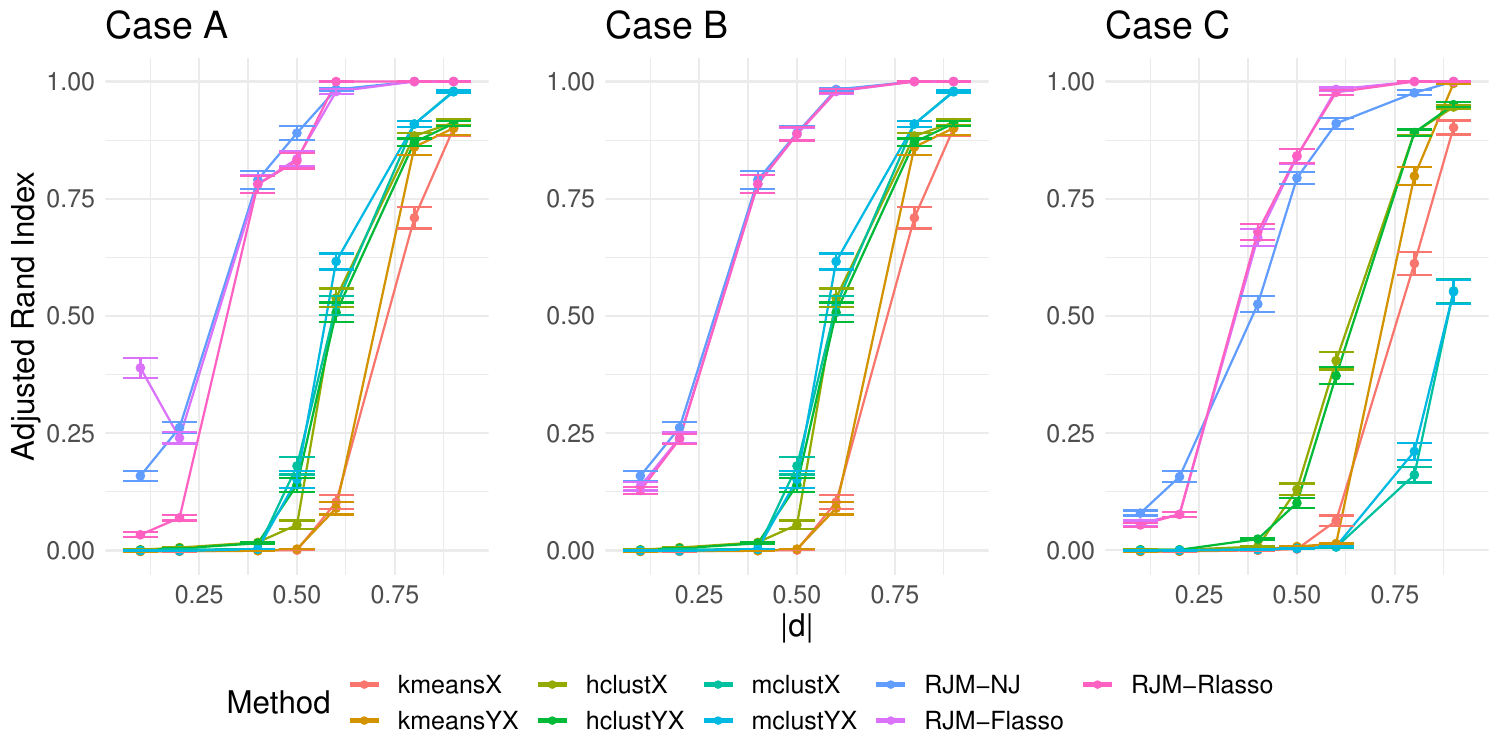}
	\caption{Second simulation, $p=250$, group assignment. Average adjusted Rand Index 
		as a function of the absolute distance ($|d|$) of the group-wise covariate means, for cases A (left), B (center) and C (right). [Error bars indicate standard errors from 20 repetitions.]}
	\label{rand250}
\end{figure}

\begin{figure}[H]
	\centering{}\includegraphics[height=7cm,width=\textwidth]{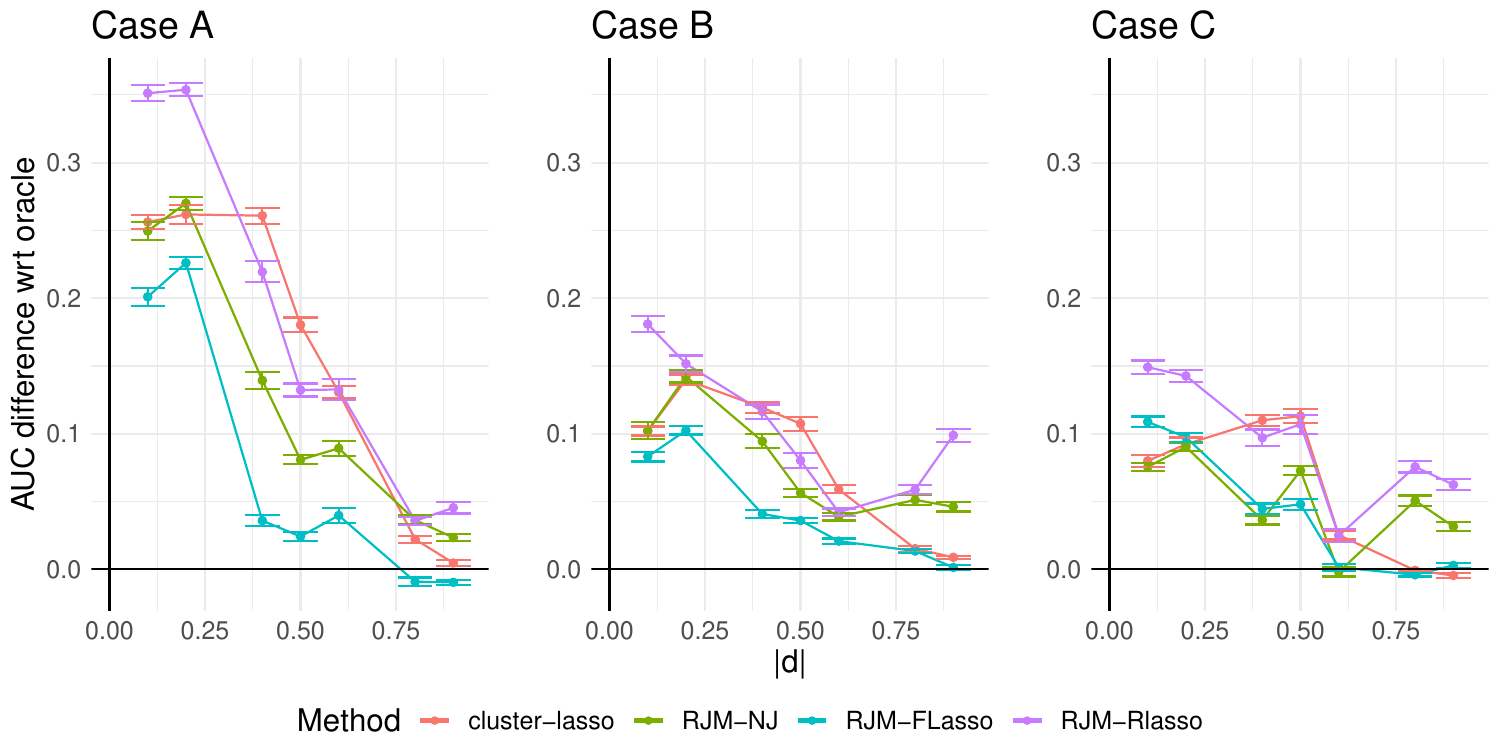}
	\caption{Second simulation, $p=100$, variable selection. AUC loss 
		from  oracle-lasso 
		as a function of  the absolute distance ($|d|$) of the group-wise covariate means, for cases A (left), B (center) and C (right). [Error bars indicate standard errors from 20 repetitions.]}
	\label{auc100}
\end{figure}

\begin{figure}[H]
	\centering{}\includegraphics[height=7cm,width=\textwidth]{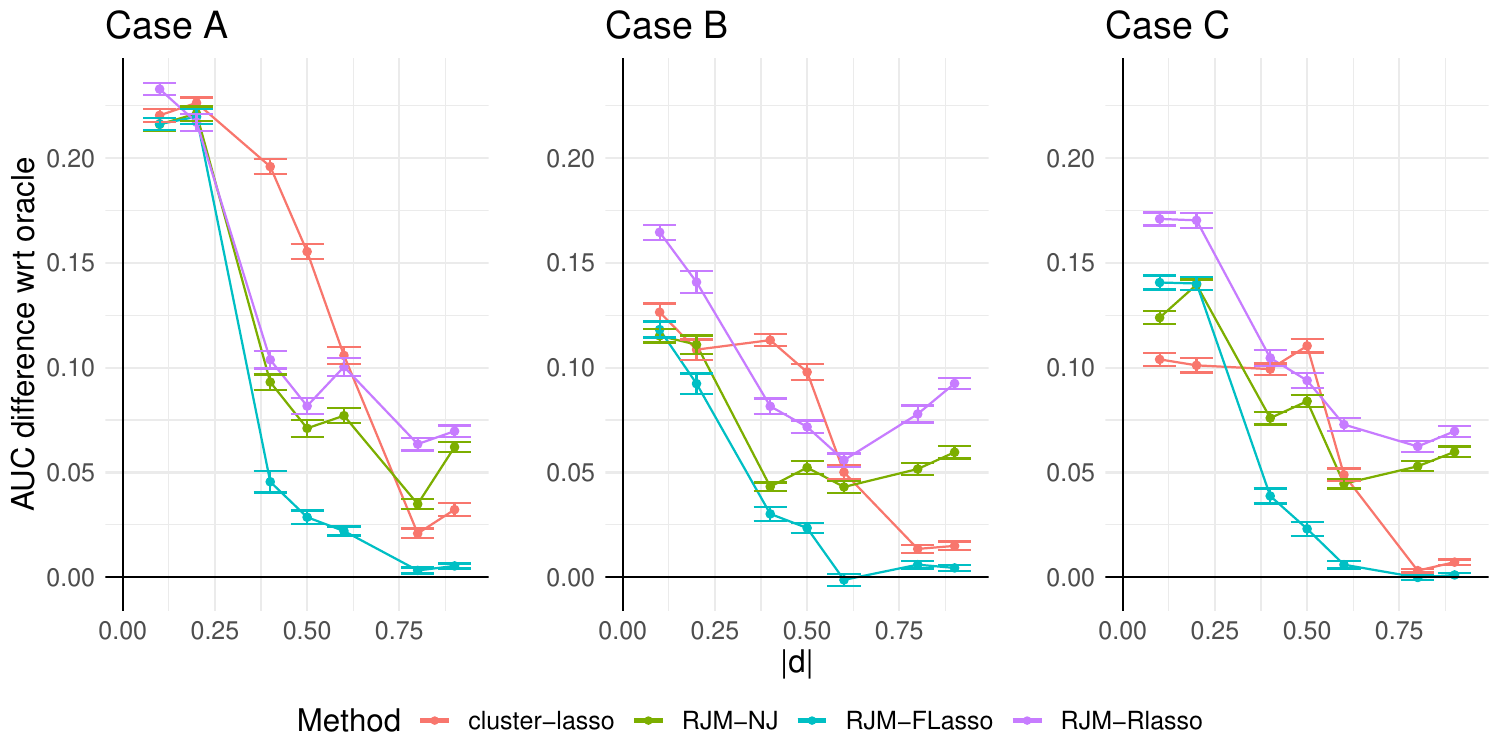}
	\caption{Second simulation, $p=250$, variable selection. AUC loss 
		from  oracle-lasso 
		as a function of  the absolute distance ($|d|$) of the group-wise covariate means, for cases A (left), B (center) and C (right). [Error bars indicate standard errors from 20 repetitions.]}
	\label{auc250}
\end{figure}

\begin{figure}[H]
	\centering{}\includegraphics[height=7.5cm,width=\textwidth]{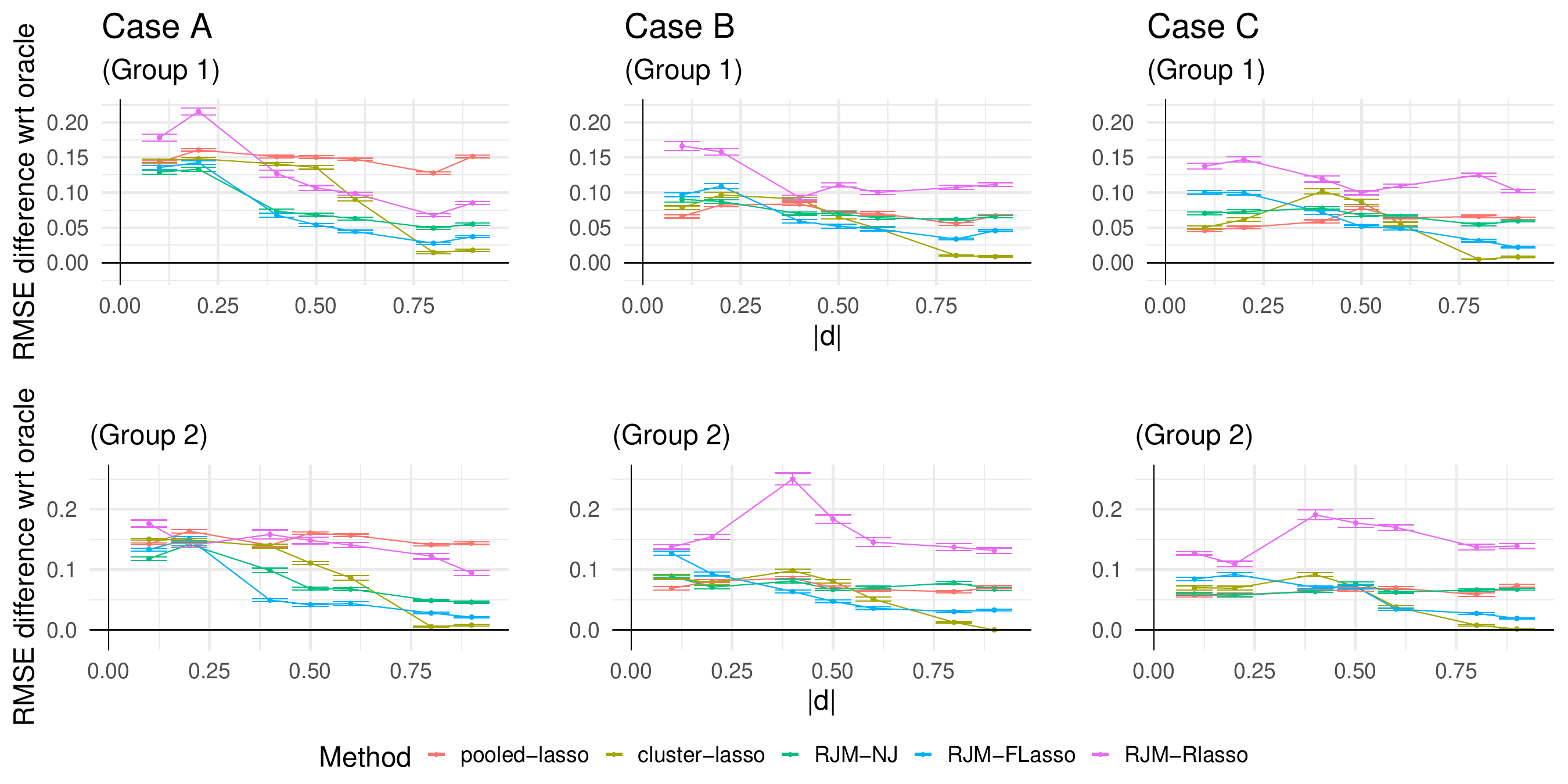}
	\caption{Second simulation, $p=100$, regression coefficients estimation. Increase in RMSE relative to the oracle-lasso 
		as a function of the absolute distance ($|d|$) of the group-wise covariate means, under group one (top) and group two (bottom), and for cases A (left), B (center) and C (right). [Error bars indicate standard errors from 20 repetitions.]}
	\label{mse100}
\end{figure}

\begin{figure}[H]
	\centering{}\includegraphics[height=7.5cm,width=\textwidth]{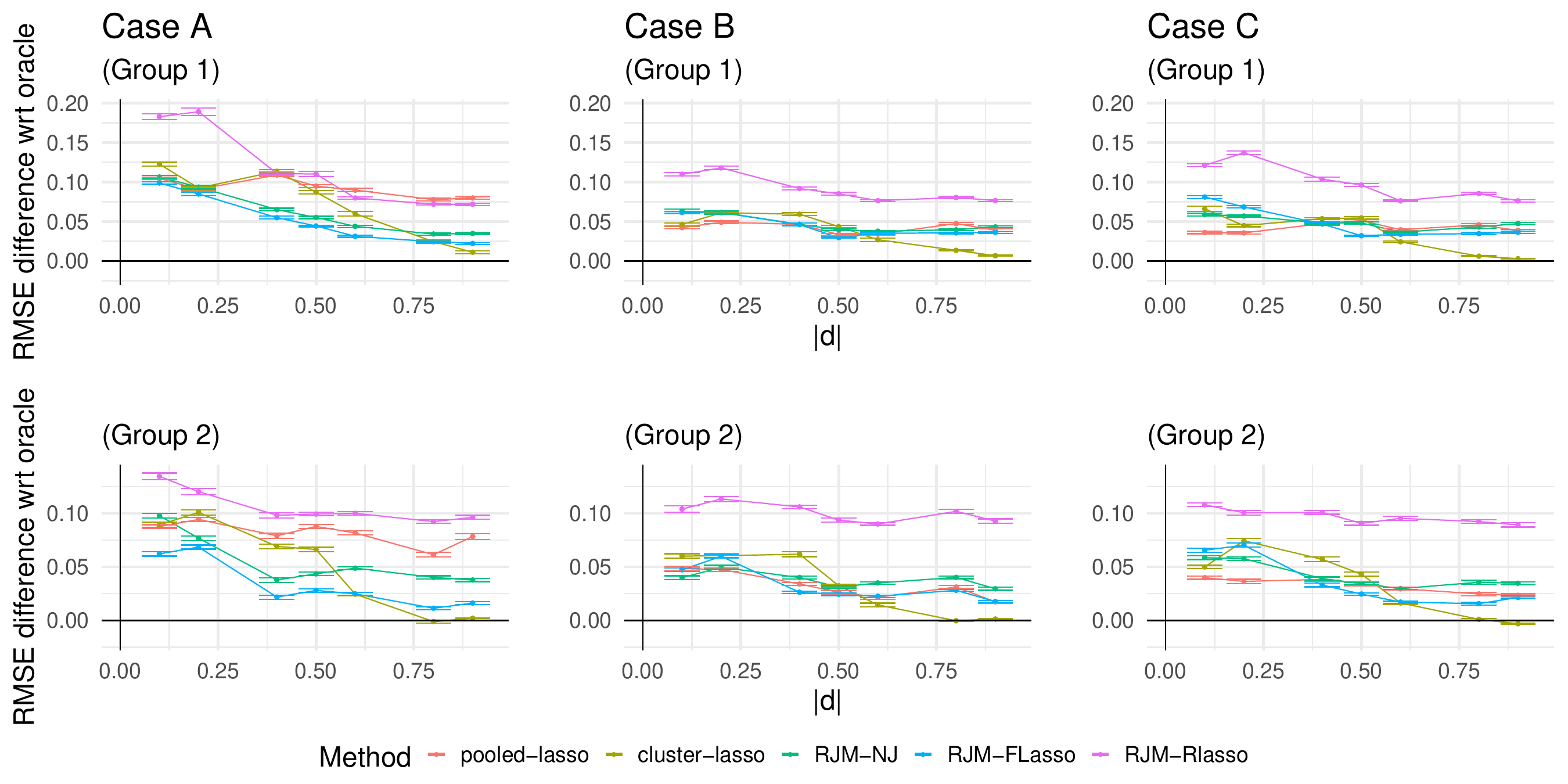}
	\caption{Second simulation, $p=250$, regression coefficients estimation. Increase in RMSE relative to the oracle-lasso 
		as a function of the absolute distance ($|d|$) of the group-wise covariate means, under group one (top) and group two (bottom), and for cases A (left), B (center) and C (right). [Error bars indicate standard errors from 20 repetitions.]}
	\label{mse250}
\end{figure}

\subsection*{Appendix H. Selection of number of groups in Section 5.2}
\label{sim3}

Here we consider all four cancer types (BRCA, KIRC, LUAD, THCA) and provide some results on cluster selection under unknown number of groups using the predictive approach described in Section \ref{prediction}.
We consider three simulation settings where the respective true number of groups is $g^*=2$ (using the BRCA and KIRC cancer types), $g^*=3$ (BRCA, KIRC, LUAD) and $g^*=4$ (further including THCA). For each setting we fit RJM models with two, three and four components.
The simulations are along the lines of Section \ref{sim2} 
considering Case A of Table \ref{sim2_param} 
for $p=100$. The conditions outlined in Table \ref{sim2_param} for the $\beta^*_j$'s at the common locations are used again for $g^*\in\{3,4\}$.
Here we use the real group-sample proportions as they occur in the TCGA data set
and assume that sample size grows with number of groups (for the simulations to be on an equal basis).
Specifically, we set $n=250\times g^*$. The resulting group sample sizes for $g^*\in\{2,3,4\}$ are $(n_1=335,n_2=165)$, $(n_1=382,n_2=188,n_3=180)$ and $(n_1=410,n_2=200,n_3=200,n_4=190)$, respectively. 
We use 80\% of the samples for training and 20\% for testing. 

\begin{figure}[h]
	\centering{}\includegraphics[height=8cm,width=\textwidth]{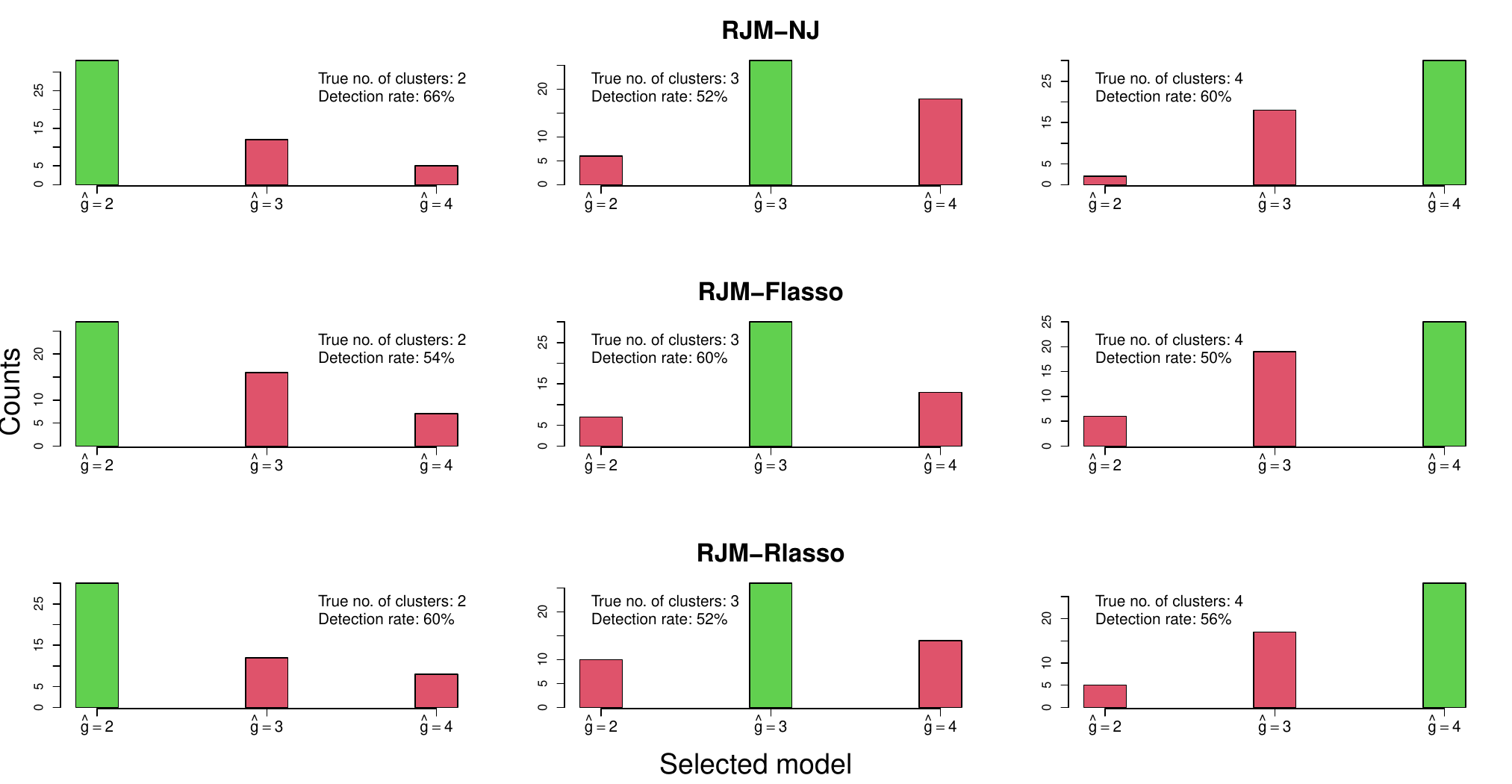}
	\caption{Second simulation, cluster selection. Barplots of selected clusters (50 repetitions) using the predictive approach in Section \ref{prediction}. Correct identification is highlighted in green; true number of clusters and detection rate of the correct model is annotated in each panel.}
	\label{selection}
\end{figure}

Figure \ref{selection} shows barplots of the selected number of clusters resulting from 50 repetitions of the simulations. As seen, the correct model is selected in majority under all cases.
The detection rate of the correct model is also annotated in each panel of Figure \ref{selection}. Here, RJM-NJ is slightly better with average overall detection rate of 59\%, while RLasso and FLasso have 56\% and 55\%, respectively.

\subsection*{Appendix I. Selection of number of groups in Section 5.3}
\label{AppI}
Here we present additional results concerning performance including a model selection step. The setting is as in Section \ref{sim4} in the main text except that sample size is equal to 200. In particular, treating in turn each of the genes as response, with all others considered as features. A model selection step is included over the number of clusters, selecting between models with $g\in\{2,3,4\}$ components based on BIC.
Here we compare with GMMs (\texttt{mclust}) and an MoE implementation from package \texttt{flexmix} \citep{flexmix}. Under the latter method we use elastic-net regularization \citep{zou_hastie_2005} in the expert networks and intercept-only multinomial gating functions (the latter is better than allowing 99 predictors to enter the gating functions in the absence of regularization for this part of the model). We note that results from \texttt{MoEclust} are not presented here as this method (based on incremental forward model search) never selected the correct model with four clusters. Results from the 100 attempts are summarized in Table \ref{cluster_sel}. 

\begin{table}[h]
	\centering
	\caption{TCGA data application, performance including a model selection step. Number of times, out of 100 applications to the TGCA data, that the methods selected two, three and four clusters based on BIC. Each time a different gene expression was used as response variable with the predictor matrix containing the remaining 99 gene expressions. The correct number of clusters is four corresponding to the four cancer types included in the dataset.}
	\begin{tabular}{lccc}
		\hline
		\noalign{\vspace{0.1cm}}
		\multicolumn{4}{c}{\textbf{Methods and cluster selection}}\tabularnewline[\doublerulesep]
		\hline \\[-2ex]
		{Estimated number of clusters} & $\hat{g}=2$ & $\hat{g}=3$ & $\hat{g}=4$   \tabularnewline
		\hline  \\[-2ex]
		{GMM (\texttt{mclust})}& 5 & 91 & 4  \tabularnewline
		{MoE (\texttt{flexmix})} &  18 & 67 & 15 \tabularnewline
		{RJM-NJ}&  17 & 18 & 65 \tabularnewline
		{RJM-FL}&  16 & 17 & 67 \tabularnewline
		{RJM-RL}&  17 & 19 & 64 \tabularnewline
		\hline
	\end{tabular}
	\label{cluster_sel}
\end{table}

\bibliographystyle{natbib}

\bibliography{biblio_new}

\end{document}